\documentclass[12pt]{iopart}  
\usepackage{graphicx}%
\usepackage{bm}%
\usepackage{bbold}
\usepackage[normalem]{ulem}
\usepackage[colorlinks=true,linkcolor=blue,citecolor=blue,urlcolor=blue]{hyperref}
\usepackage{xcolor}
\usepackage[margin=1.5cm]{geometry}
\usepackage{cite}
\usepackage{pstricks,pst-node}
\usepackage{amsfonts}

\begin{document}

\title{Geometric properties of qudit systems}

\author{J.A. L\'opez--Sald\'ivar{$^{1,2}$}, O. Casta\~nos{$^{3,\dag}$}, S. Cordero{$^3$}, E. {Nahmad--Achar}{$^3$}\footnote{Author to whom any correspondence should be addressed} and R. {L\'opez--Pe\~na}{$^3$}}
\address{
$^1$
Russian Quantum Center, Skolkovo, Moscow 143025, Russia
}
\address{$^2$
National University of Science and Technology “MISIS”, Moscow 119049, Russia
}
\address{%
{$^3$} Instituto de Ciencias Nucleares, Universidad Nacional Aut\'onoma de M\'exico, Apartado Postal 70-543, 04510  Mexico City, Mexico}
\address{
{$^\dag$} On sabbatical leave at University of Granada, Spain
}

\ead{\mailto{julio.lopez.8303@gmail.com}, \mailto{ocasta@nucleares.unam.mx}, \mailto{sergio.cordero@nucleares.unam.mx}, \mailto{nahmad@nucleares.unam.mx}, \mailto{lopez@nucleares.unam.mx}}

\date{\today}

\maketitle

\newcommand{\op}[1]{{\bm{#1}}}
\newcommand{\bra}{\langle}
\newcommand{\ket}{\rangle}
\newcommand{\diag}[1]{{\rm DiagM}\,\left(#1\right)}
\newcommand{\diagb}[1]{{\rm Diag}\,\left(#1\right)}
\newcommand{\new}[1]{\textcolor[rgb]{0,0,0.7}{#1}}
\newcommand{\old}[1]{\textcolor[rgb]{0.5,0.5,0.5}{\sout{#1}}}
\newcommand{\nota}[1]{\textcolor[rgb]{1,0,0}{{#1}}}
\newcommand{\newE}[1]{\textcolor[rgb]{0,0.8,0}{#1}}

\section*{Abstract}
We discuss in general how to geometrically visualize a qudit system, with a particular interest in thermal states.
The principle of maximum entropy is used to study the geometric properties of an ensemble of finite dimensional Hamiltonian systems with known average energy. These geometric characterizations are given in terms of the generalized diagonal Bloch vectors and the invariants of the special unitary group in $n$ dimensions. As examples, Hamiltonians written in terms of linear and quadratic generators of the angular momentum algebra are considered with $J= 1$ and $J=3/2$. For these cases, paths as functions of the temperature are established in the corresponding simplex representations, which show first- and second-order quantum phase transitions, as well as the adiabatic evolution of the interaction strengths (control parameters) of the Hamiltonian models. For the Lipkin-Meshkov-Glick Hamiltonian the quantum phase diagram is explicitly shown for different temperature values in parameter space.
\vspace{0.1 in}

\noindent {\bf Keywords}: {\it thermal qudit states, geometric properties, maximum entropy, invariant spaces, LMG model}

\section{Introduction}

Quantum information theory has become an important of study in physical systems, mainly because the processing of quantum information has yielded a deep understanding about the fundamental aspects of quantum mechanics. The purpose of this field is the exploitation of the quantum features with a technological purpose~\cite{bennett98, nielsen11, werner01, zoller05}. Differences between classical and quantum information are related to the possibility of having secure communications, to the solution of certain mathematical problems, such as the efficient  lookup of data in a generic disordered database, cryptography, dense coding, teleportation in discrete and continuous variables, counterfit impossibility, and to having a quantum computer~\cite{lapierre21, horodecki22, bengtsson17, bruss19}.

The set of all pure and mixed physical finite quantum states is given by the Hermitian matrices with unit trace and having non-negative eigenvalues, the so-called {\it density matrices}. Linear and von Neumann entropies, calculated through reduced density matrices, measure entanglement in bipartite pure systems; thus, density matrices do not only provide a means to study the phase diagrams of interacting systems, but also their quantum correlations. There are several parametrizations of the density matrix for finite $n$-dimensional systems: The generalized Gell-Mann vectors, those associated to the generators of the special unitary algebras $\mathfrak{su}(n)$, the polarization operators (spherical tensors), the Casimir invariants of SU($n$), and the canonical coset decomposition of unitary matrices~\cite{fano57, byrd03, kimura03, akhtarshenas07, bertlmann08}. A parametrization and representation of states using a decomposition based on Gell-Mann matrices is given in~\cite{kurzynski16, sarbicki12}.  An interesting feature of unitary group representations for quantum systems is their correspondence with different projective geometries, as the Fano projective plane \cite{Grunbaum2009-ga} with the $2$-simplex of the qutrit system, or the projective space PG(3,2) \cite{Hirschfeld1986-uz} with the $3$-simplex defined for a four-level system. See~\cite{rau21} and references therein for a study and historical perspective of  the connection between quantum information processes to projective geometries and combinatorial design.

Qudits systems can enhance the performance in quantum computing because they provide a larger state space to store and process information, and thus increase the variety of quantum circuits that can be devised to solve specific problems~\cite{wang20,mandilara24}. The qutrit, an eight-dimensional convex body, can be visualized through the set of all possible two-dimensional cross-sections. Its geometric structure has been characterized using symmetric informationally complete measurements~\cite{tabia13}, and different visualization manners for a qutrit exist~\cite{kurzynski16, xie20, mandilara24}.  Recently, in the probability representation, the simplex associated to the reduced density matrix for one-particle was obtained for a system of $N_A$ atoms interacting dipolarly with a two-mode electromagnetic field, for the atomic configurations $\Xi, \Lambda$ and $V$~\cite{cordero21, lopez-pena21}.

It is instructive, therefore, to study the geometric properties of density matrices in general. Since they have unit trace, their diagonal elements correspond to probabilities of finding the quantum state in one or another element of the Hilbert space basis, upon measurement. This property lends itself to study the geometric properties of density matrices in $n$-dimensional simplexes, where $n+1$ is the dimension of the system's Hilbert space. Being Hermitian means that physical states, in contrast with purely mathematical states, will occupy only certain regions of the simplex; further restrictions on the simplex region to be occupied are given by the positivity condition of the characteristic polynomial of the density matrix, thus the study of these regions in different representations constitute an excellent aid in understanding the properties of the system. Maximally mixed states, ordinary mixed states, and pure states, will occupy different regions in the simplex, and certain curves on the simplex will correspond to the evolution of one state into another. Paths as functions of the temperature may be established for thermal states, as well as adiabatic evolutions.

In this contribution, the physical space associated to the diagonal generalized Bloch vectors, the probability representation, and the Casimir invariants is determined, in order to describe the well-defined density matrices of systems of qubits, qutrits, and ququarts~\cite{goyal16}. In particular, we will consider in this work the generalized Bloch vectors because they allow to determine the density matrix on the basis of actual measurements (cf.~\ref{ap1}). The Principle of Maximum Entropy (PME) is used to determine the density matrices of ensembles of qudits with a defined average energy. These ensembles are described by Hamiltonians given in terms of linear and quadratic generators of the angular momentum algebra. By means of the diagonal decomposition of the density matrices we may visualize geometrically the purity properties of ensembles with known average energy. In particular, the available paths as a function of the temperature in the corresponding simplex representations are established for ensembles of qutrits and ququarts.

Section~\ref{s.VM} gives a description of the Principle of Maximum Entropy for an ensemble of particles with a fixed average energy, which yields the Boltzmann probability density and at the same time minimizes the Helmholtz free energy of the quantum system. Different matrix representations of a density operator of an $n$ dimensional quantum system are also given, these being the probability, the Bloch-Gell-Mann, and the invariant cases.  
In addition, a general discussion of the thermal trajectories as a function of the temperature is given, from $T\to 0$ (pure state) to $T\to\infty$ (the most mixed state).
In section~\ref{s.dqtqq} the diagonal representations for the qutrit and ququart systems are established together with the corresponding mapping between them. One is then able to exhibit the localization of the physical density matrices in the probability ($p$)-, Gell-Mann ($\lambda$)-, and invariant ($t$)-spaces,
with and without eigenvalue degeneracy, the latter of which
is associated to the higher dimensional orbits of the qutrit and ququart systems. In section~\ref{s-thermal}, linear and quadratic Hamiltonian models in terms of the angular momentum operators are considered, particularly for $J=1$ (a realization in terms of identical but distinguishable dimers) and $J=3/2$ (a realization in terms of identical but distinguishable trimers). The behavior of the thermal states for the qutrit and ququart is obtained and visualized in the $p$-, $\lambda$- and $t$-spaces. For the Lipkin-Meshkov-Glick (LMG) Hamiltonian model particular attention is paid to the quantum phase transitions and their influence on the characteristics of the corresponding thermal states in the diagonal representation. Section~\ref{s-conclusions} gives a summary of the main results and additional remarks.

\section{Variational thermal density matrix. }
\label{s.VM}

The first explicit relation made between physical entropy and information is associated to the relation between energy and information. This goes back to Maxwell's demon paradox introduced in 1867, which implies an apparent violation of the second law of thermodynamics. The connection of information theory to physics helps to solve Maxwell's demon paradox with Landauer's principle, which enables one to extract energy from information~\cite{benenti07,grandy97}. Shannon~\cite{shannon48} establishes that in a message of $n$ symbols, having transmission probabilities $(p_1,p_2,\cdots,p_n)$, the amount of information is
\begin{equation*}
\mathcal{H}_S = -  \sum^n_{i=1} p_i \ln p_i \, .
\end{equation*}

E. Jaynes in 1957 noted that Shannon had actually uncovered a fundamental element of probability theory, viz., $\mathcal{H}_S \equiv S$
defines the entropy of a probability distribution on an exhaustive set of mutually exclusive alternatives. The PME establishes that
{\it the distribution $\{ p_k\}$ that maximizes $S$ subject to the constraints imposed by the available information is the least biased description of what we know about the set of alternatives}~\cite{jaynes57a,jaynes57b, louisell90}. It has been proven that the PME is a correct method of inference when new information is given in terms of expectation values~\cite{shore80}.

The PME provides a variational procedure for constructing prior probabilities on a given evidence. If we do not know something about a system, this rule tells us that the system has the same probability of staying in any of all its possible states, i.e., $p_\ell=1/n$ and $S=\ln n$. If the system is in one particular state $k$, one has $S=0$. If the expectation value of the energy is known, the PME determines the Gibbs canonical ensemble, if the energy and number of moles are known one gets the Gibbs grand canonical ensemble, if the expectation values of the energy and angular momentum are known the PME determines the Gibbs rotational ensemble~\cite{jaynes62}.  Additionally, the use of the PME for physical systems in thermal equilibrium can be extended to solve non-equilibrium problems because it can be applied to any physical quantity, be it density of particles, density of kinetic energy, components of the stress tensor, intensity of the magnetization, and so on.

In quantum statistical mechanics the entropy is defined in terms of the density matrix
\begin{equation}
{S} = -k_B {\rm Tr}({\op{\rho} \ln{\op{\rho}}}) \, ,
\end{equation}
with the constraints ${\rm Tr}{\op{\rho}}=1$ and ${\rm Tr}{\op{\rho}^2}\leq1$. If one considers finite $n$-dimensional systems, the density operator has also another restriction: ${\rm Tr}{\op{\rho}^2}\geq 1/n$. 
According to the PME, {\it prior} knowledge of an ensemble of quantum systems can be used to maximize the entropy. If the average energy ${\rm Tr}( \op{\rho} \, \op{H})= \bar{E}$ is known, one must then take the following variations: $\delta S=0$, $\delta({\rm Tr} \op{\rho} -1)=0$, and $\delta({\rm Tr}( \op{\rho} \, H)- \bar{E}) = 0$, implying that~\cite{reichl16}
\begin{equation}
{\rm Tr}\left[ (1 + \ln \op{\rho} +\epsilon + \beta \, \op{H}) \,\delta\op{\rho}\,\right] =0 \, ,
\end{equation}
where we have introduced two Lagrange multipliers $\epsilon$ and $\beta$. From the last expression one obtains the density matrix $\op\rho$ and, taking its trace, $\epsilon$ is determined, yielding
\begin{equation}
\op\rho = \frac{1}{Z} \exp{(-\beta \, \op H)} \, ,
\label{therm_gen}
\end{equation}
with the partition function $Z={\rm Tr}\left[\exp{(-\beta \op H)}\right]$.
In order to determine $\beta$, one considers a cavity filled with photons in thermal equilibrium with the walls at temperature $T$. In this case, and by means of the {\it Correspondence Principle}, one obtains $\beta=\frac{1}{k_B\, T}$.
The maximization requirement of the entropy yields the expression
\begin{equation}
T \, S - U + F =0 \, ,
\end{equation}
with the identification of the internal energy of the system $U=\langle \op H \rangle$ and the Helmholtz free energy $F = -k_B T \ln Z$. Therefore the PME implies also a minimum Helmholtz free energy for the system.

Expression~(\ref{therm_gen}) may be obtained by considering a closed system constituted by the Hamiltonian associated to the physical system of interest and the Hamiltonian of a heat bath representing the environment, in which the interaction term can be ignored. In this case the composite system can be described by a micro-canonical ensemble in the energy interval $E$ and $E+ \delta E$ with $\delta E\ll E$. If the number of degrees of freedom of the bath is very large the reduced density matrix takes the form given in Eq.~(\ref{therm_gen}). This procedure has led to the answer of when is it possible to represent the reduced density matrix of the system $\op \rho_S$ in terms of a wave function, $\psi$ with $\op \rho_\psi=\op \rho_S$~\cite{linden02, goldstein06, tasaki98}.

Note, then, that the variational method maximizes the von Neumann entropy and simultaneously minimizes the energy.

\subsection{Matrix representations.}
\label{ss.matrixR}

In order to study finite mixed quantum systems the use of the density matrix formalism is necessary. Besides ({\it vide infra}), to consider its diagonal form is convenient if one wishes to classify the different types of density matrices according to the dimension of the orbits in the tangent space.

The set of diagonal density matrices of dimension $n$ can be denoted in terms of their eigenvalues,
\begin{equation}
\op D_n = \diag{ p_1,p_2, \cdots p_{n-1}, p_n } \, ,
\end{equation}
with $0\leq p_k\leq 1$ and $\sum^n_{k=1} p_k=1$. 
%
This set then forms a simplex in $n-1$ dimensions (cf. e.g.~\cite{coxeter48}). A general density matrix can be constructed from them via
\begin{equation}
\op \rho = \op{U} \, \op D_n \op{U}^\dagger \, ,
\end{equation}
where $\op{U}$ denotes a unitary transformation in $n$ dimensions, U$(n)$. 
The set of transformations leaving $\op D_n$ invariant form a group called the stability group ${\cal H}$, i.e.,
\begin{equation}
\op{T}_{\cal H} \, \op D_n \, \op{T}_{\mathcal{H}}^\dagger = \op D_n \quad \rightarrow \quad \op \rho = \op{\Omega} \, \op D_n \op{\Omega}^\dagger \, ,
\end{equation}
with $\op{\Omega}$ being an element of the quotient group U$(n)/{\cal H}$ and the unitary transformation can be decomposed as $\op{U} = \op{\Omega} \, \op{T}_{\cal H}$. This result allows us to characterize the $n \times n$ density matrices by the orbit of a point in $\op D_n$ under the action of the quotient group U$(n)/{\cal H}$. The characterization can be done by denoting the degeneracy of the eigenvalues $p_j$ by
\begin{equation}
{\cal M} =\{m_1,m_2, \cdots, m_{\ell-1}, m_\ell \} \, .
\end{equation}
Then the stability group is defined by the tensorial product
\begin{equation}
{\cal H} = {\rm U}(m_1) \otimes {\rm U}(m_2) \otimes, \cdots \otimes {\rm U}(m_\ell) \, .
\end{equation}
The quotient group is called a complex flag manifold denoted by
\begin{equation}
{\cal F} = {\rm U}(n)/({\rm U}(m_1) \otimes, \cdots, \otimes {\rm U}(m_\ell)) \, .
\end{equation}
These mathematical results~\cite{gilmore05} are going to be applied for the particular Hamiltonian systems considered here, viz., the linear and quadratic Hamiltonians in terms of the angular momentum generators.

The generators of an $n$ dimensional unitary group can be characterized by the operators $ \op{A}_{kj}=|k\ket\bra j|$ with $j,k=1,2,\cdots,n$, and through them any unitary matrix can be constructed, 
\begin{equation}
(\op{\rho})_{jk} := \bra j|\op{\rho}|k\ket={\rm Tr}\,[\op{\rho}\op{A}_{kj}]\,
\end{equation}
where $\op{A}_{kj}\op{A}_{\ell m} = \delta_{j\ell}\op{A}_{km}$. The positive coefficient $p_j:=(\op{\rho})_{jj}$ denotes the probability of finding the system in the state given by the projector $\op{P}_{j}:=\op{A}_{jj}$, while the non-diagonal elements $a_{jk}:=(\op{\rho})_{jk}$ stand for the probability amplitude for the transition $|k\ket\to|j\ket$. Then, in general, a density matrix can be written as
\begin{equation}\label{eq.rhoG}
\op{\rho} = \sum_{j=1}^n p_j\,\op{P}_{j} + \sum_{j\neq k}^n a_{jk}\,\op{A}_{jk}\,;\qquad a_{kj}=a_{jk}^*\ \land \ {\rm Tr}\,[\op{\rho}]=1\,,
\end{equation}
together with the positivity condition. Note that, if the non-diagonal terms satisfy $a_{jk}=\sqrt{p_jp_k}\,e^{i(\theta_k-\theta_j)}$ then $\op{\rho}$ represents a pure state: ${\rm Tr}\,(\op{\rho}^2)=1$.
\begin{enumerate}
\item {\it Probability representation}. The simplest representation is to consider the set of  probability values for a fixed set of projectors $\op{P}_j$, i.e., $p_j={\rm Tr}\,[\op{\rho}\op{P}_j]$, which defines a probability-space referred to hereafter as the {\it $p$-space}. In particular, if the density matrix $\op{\rho}$ is given in terms of its eigen-projectors, i.e., when $\op{\rho}\op{P}_j = p_j\op{P}_j$ for each $\op{P}_j$, the matrix is written in a diagonal form as 
\begin{equation}\label{eq.rhoP}
\op{\rho} = \sum_{j=1}^n p_j\,\op{P}_j:=\diag{p_1,\,p_2,\,\dots,\,p_n}\,;\qquad \op{P}_j\op{P}_k=\delta_{jk}\op{P}_j\,.
\end{equation}
This set of projectors defines a subset of density matrices which conmute with each other $[\op{\rho},\op{\rho}']=0$. 

In this representation, the set of all qudit states can be represented by a ball in $d=n$ dimensions, in terms of which the corresponding simplex may be written in terms of the probabilities
\begin{equation}
\vec{p} = \vec{p}_e + \frac{\vert \vec{\lambda}\vert}{\sqrt{2}}\,\hat{e}_{r} = \vec{p}_e + \frac{r}{\sqrt{2}}\,\hat{e}_{r} \,,
\end{equation}
where $\vec{p}_e = \frac{1}{n} (1,1,\ldots,1)$ is the most mixed state, $\vert \vec{\lambda}\vert$ is the magnitude of the Bloch vector $\vec{\lambda} = (\lambda_1,\,\lambda_2,\,\ldots,\,\lambda_{n^2-1})$, and $\hat{e}_r$ is the unit radial vector describing the sphere in $n$-dimensions (see~\ref{ap.pvectors}).

The relation of one point in the simplex with one physical state is unique only when the diagonal matrix $\op{\rho}$ is written in the fixed set of its eigen-projectors $\op{P}_j$, because the $p$-space forgets the physical nature of the problem. 
\item {\it Gell-Mann representation}. Using the Gell-Mann operators $\op{\Lambda}_j$ the density operator can be written in the form~\cite{kimura03, bertlmann08, petruccione12}
\begin{equation}\label{eq.rhoL}
\op{\rho}= \frac{1}{n} \op{I} + \frac{1}{2} \sum_{j=1}^{n^2-1} \lambda_j\,\op{\Lambda}_j \,;\qquad {\rm Tr}\,[\op{\Lambda}_j]=0\,;\quad {\rm Tr}\,[\op{\Lambda}_j\op{\Lambda}_k]=2\delta_{jk}\,.
\end{equation}
Since $\op{\Lambda}_j$ are traceless operators, a direct comparison between equations~(\ref{eq.rhoL}) and equations (\ref{eq.rhoG}) for the diagonal representation of a density matrix, shows that one may choose $n-1$ diagonal operators, providing the $n-1$ free parameters $\vec{\lambda} = (\lambda_{k_1},\,\lambda_{k_2},\,\dots,\,\lambda_{k_{n-1}})$ defining an $(n-1)$-dimensional space (referred to as the {\it $\lambda$-space} hereafter). 
By using the diagonal operators~(\ref{eq.LambdaD}), one may establish the relationship between the free $\lambda_{k_j}$ ($k_j=n^2-n+j$) parameters and probabilities $p_s={\rm Tr}\,[\op{\rho}\op{A}_{ss}]$ as 
\begin{equation*}
p_s = \frac{1}{n} + \frac{1}{2} \sum_{\ell=1}^{n-1} a_{k_\ell s}\, \lambda_{k_\ell}\,, \quad  a_{k_\ell s}:=(\op{\Lambda}_{k_\ell})_{ss}\,, \quad s=1,\,2,\,\dots\, n\,.
\end{equation*}
The set of values $a_{\ell s}$ and the constriction $0\leq p_s\leq1$ yield the set of aceptable values $\vec{\lambda}$ in $\lambda$-space for which the matrix $\op{\rho}$ represents a physical state of the system. Notice that the relationship between probabilities and the free parameters $\vec{\lambda}$ is linear, so each point in $p$-space is mapped to one and only one point in $\lambda$-space, and hence the transformation is invertible, i.e., the map
\begin{equation}\label{eq.pML}
\vec{p} = \op{M}\vec{\Lambda}\,; \quad \vec{\Lambda}:=(\lambda_{k_1},\,\lambda_{k_2},\,\dots,\,\lambda_{k_{n-1}},1)\,,
\end{equation}
with
\[\op{M} =\frac{1}{2}\left(\begin{array}{c c c c c} 
a_{k_11} & a_{k_21} &\dots &a_{k_{n-1}1} & 2/n \\
a_{k_12} &a_{k_22} & \dots & a_{k_{n-1}2}& 2/n \\
\vdots &\vdots &\ddots & \vdots & \vdots \\
a_{k_1n} &a_{k_1n} & \dots& a_{k_{n-1}n}& 2/n
\end{array}\right)\,,\]
has inverse matrix $\op{M}^{-1}$. Clearly, the zero vector $\vec{\lambda}_e$ or $\vec{\Lambda}_e=(0,\,0,\dots,\,0,\,1)$ represents the most mixed state; while the set of pure states are given by the set of vectors $\vec{\Lambda}_j$ for which  $\vec{p}_j=\op{M}\vec{\Lambda}_j$, and each one of these corresponds to a vertex in $\lambda$-space.
\item {\it Invariants representation}. An alternative form to characterize a set of density matrices of dimension $n$ is via the named invariants $t_s:={\rm Tr}\, [\op{\rho}^s]$ for $s=1,\,2\,\ldots$ Besides $t_1=1$, it is a fact that there are $n-1$ invariants which are independent; typically, the first ones are considered, i.e.,  
\begin{equation}\label{eq.invariants}
t_\ell := {\rm Tr}\, [\op{\rho}^\ell]\,; \qquad \ell=1,\,2,\,,\dots,\,n\,.
\end{equation}
This means that any invariant $t_s$ for $s>n$ can be given in terms of $t_1,\,t_2,\,\dots,\,t_n$. Since $t_1=1$ is a constant, the set of invariants defines  an $(n-1)$-dimensional space, the {\it invariants-space}, hereafter referred to as {\it $t$-space}, with points of the form $\vec{t}=(t_2,\,t_3,\,\dots,\,t_n)$, each delimited by the condition
\[\frac{1}{n^{\ell-1}}\leq t_\ell\leq 1\qquad \ell=2,\,3,\,\dots,\,n\,,\]
where the lower and upper bounds are calculated for the most mixed and pure states, respectively. 
Similar to the simplex in $p$-space or $\lambda$-space, the region of $t$-space for which each point defines a physical state of the system has $n$ vertices, given by 
\begin{equation}\label{eq.vertices.t}
\vec{t}_k:=\left(\frac{1}{k},\,\frac{1}{k^2},\,\dots,\,\frac{1}{k^{n-1}}\right)\,;\qquad k=1,\,2,\,\dots,\,n\,.
\end{equation}
The points $\vec{t}_1$ and $\vec{t}_n$ correspond to the pure and most mixed state, respectively. The other ones, $\vec{t}_k$ with $1\neq k\neq n$, correspond to the $k$-maximum mixed states, i.e., states composed with only $k$ pure states. Thus, for a fixed set of pure states, each vertex represents 
\[N_k=\frac{n!}{k!(n-k)!}\,,\]
physical states, while any other point $\vec{t}\neq\vec{t}_k$ represents $n!$ physical states; in other words, one may map the simplex to $t$-space but this map is not invertible.
\end{enumerate}

One may study and visualize the behavior of any finite dimensional density matrix in each of the matrix representations considered in this section, that is, in the $p$-, $\lambda$-, and $t$- spaces. Examples of interest can be the following: (i) the temporal unitary evolution of a mixed state constituted by a linear combination of energy eigenstates, multiplied by coefficients $p_k$ satisfying $p_k \geq 0$ and $\sum_k p_k=1$; (ii) the adiabatic evolution of the parameters appearing in a Hamiltonian operator and yielding quantum phase transitions; and (iii) the movement in the mentioned spaces when the temperature between the system and the bath is changing. (iv) Finally, the use of a master equation to determine a non-unitary temporal evolution of a quantum system may also be studied.

\subsection{Thermal density matrix at endpoints.}

In this work, we are interested in studying the case of Hamiltonians with a finite number of accesible energy levels; in this case one has two limit points, viz. $\beta\to0$ (infinite temperature) and $\beta\to\infty$ (zero temperature). Without loss of generality, one may consider the case of $n$-level systems and that the energy levels have been labeled increasingly, i.e.,
\[h_1\leq h_2\leq\cdots  \leq h_n\,.\]
The thermal matrix density matrix Eq.~(\ref{therm_gen}) in the Hamiltonian basis is a diagonal matrix
\begin{equation*}
\op{\rho}_D = \diag{p_1,\,p_2\,,\dots,\,p_n}\,;\qquad p_j = \frac{e^{-\beta h_j}}{\sum_{k=1}^ne^{-\beta h_k}}\,,
\end{equation*}
and in the limits one has:
\begin{enumerate}
\item Case $\beta\to0$: in this case all values $p_j$ go to a constant
\[\lim_{\beta\to 0} p_j\to\frac{1}{n}\,,\]
representing the most mixed state, where any eigenstate of the system is accesible with the same probability.
\item Case $\beta\to\infty$: for this case any value $p_j$ corresponding to an excited state ($h_j>h_1$) vanishes, while for states with energy equal to the ground state energy ($h_j=h_1$) one has 
\[\lim_{\beta\to \infty} p_j\to\frac{1}{k}\,,\]
where $k$ stands for the degeneracy of the ground state of the system. In this sense, the density matrix becomes a particular ground state projector, with equal probability for each fixed ground state in the basis, as expected because Eq.~(\ref{therm_gen}) is the density matrix for the ground state with maximum entropy for finite temperature.
\end{enumerate}
The endpoints of the thermal state correspond to the vertices Eq.~(\ref{eq.vertices.t}) of the physical $t$-space in the invariants representation.

\begin{center}
\begin{figure}
\includegraphics[width=0.48\linewidth]{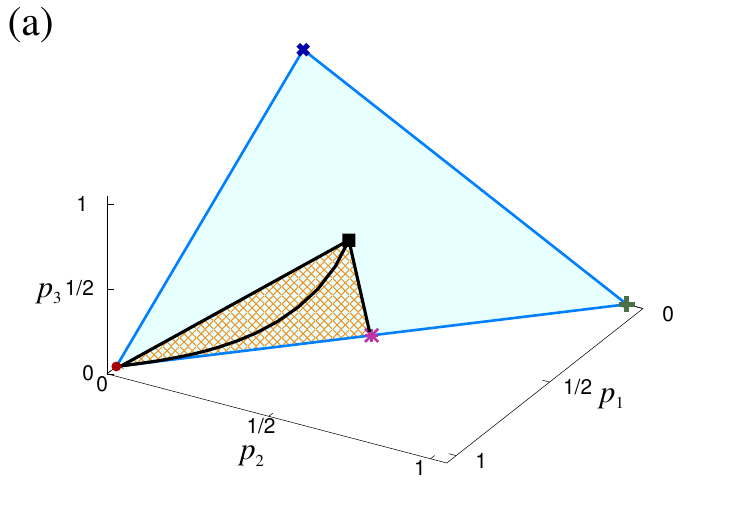}
\includegraphics[width=0.48\linewidth]{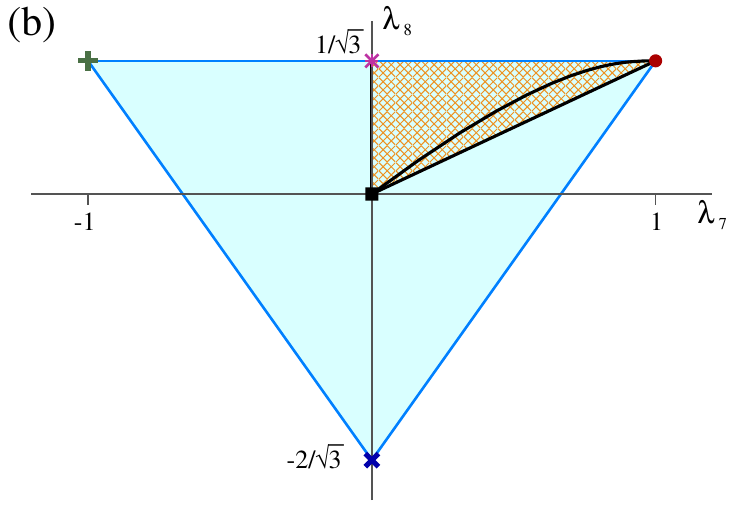}
\caption{The simplex region in $p$-space (a), and the projection from $p$-space to $\lambda$-space (b), are shown for a $3$-level system. In both cases, the vertices representing pure states are indicated by a dot-symbol (occupation probability of the ground state), plus-symbol (occupation probability of the first excited state) and times-symbol (occupation probability of the second excited state). The central square-symbol represents the most mixed state, and the asterisk-symbol marks the middle point between $p_1$ and $p_2$. Solid black lines correspond to the points of the thermal state as a function of the temperature, so the straight line from the vertex corresponds to starting from a pure ground state; the straight line which connects the central point with the middle point along the edge corresponds to the case of two degenerate occupation probabilities of the ground state; and the black curve shows the qualitative behavior for the general case of a density matrix without degenerate eigenvalues.}
\label{f.simplexPL}
\end{figure}
\end{center}

These results are independent of the matrix representation of the density matrix. So, in $p$-space the thermal density matrix as a function of the temperature moves from a vertex (or from the mid-point in the (simplex edge) subspace of degenerate ground state energy) to the simplex centroid (the most mixed state) along a trajectory (cf. Fig.~\ref{f.simplexPL} and the discussion in Subsection~\ref{s.dqtqq-1}). A similar result is obtained in $\lambda$-space for the Gell-Mann representation, while in $t$-space this trajectory journeys from the vertex $\vec{t}_k$ (with $k$ the degeneracy of ground state) to the vertex $\vec{t}_n$.  

Note that to consider the density matrix in its diagonal representation facilitates the study of its properties as functions of the set of variables involved in an $n$-dimensional quantum system.

\section{Geometric properties of the qutrit and ququart systems.}\label{s.dqtqq}

To describe the general properties of the qutrit and ququart systems, we use their corresponding diagonal representations. A qutrit system has a geometric representation in a hyperplane of a $3$-dimensional space ($p$-space) or in a $2$-dimensional space ($\lambda$- and $t$-spaces), while the ququart system requires of an hyperplane in a $4$-dimensional space ($p$-space) or a $3$-dimensional space ($\lambda$- and $t$-space). Here we review these representations for general diagonal density matrices, i.e., we suppose that the density matrix of the qutrit is given by $\op{\rho}=\diag{\rho_{11},\,\rho_{22},\,\rho_{33}}$ and for the ququart $\op{\rho}=\diag{\rho_{11},\,\rho_{22},\,\rho_{33},\,\rho_{44}}$, and point out some interesting features of their representations.

\subsection{Qutrit system}
\label{s.dqtqq-1}

In the probability representation, the numerical eigenvalues are limited by the normalization condition ${\rm Tr} \, \op{\rho}_D=1$, i.e., $p_1+p_2+p_3=1$, and by the probabilistic interpretation of the eigenvalues of the density matrix $0\leq p_j \leq 1$ ($j=1,2,3$). The spanned space from these conditions is the simplex shown in  Fig.~\ref{f.simplexPL}(a) whose vertices are located at $\vec{p}_1:=(1,0,0)$, $\vec{p}_2:=(0,1,0)$, and $\vec{p}_3:=(0,0,1)$. These three points depict each one of the pure states (${\rm Tr} \, \op{\rho}^2=1$) with one eigenvalue equal to one and the other two equal to zero. The edges delimiting the triangle are parametrized by

\begin{equation} 
\vec{p}_{jk} =\vec{p}_j +\left(\vec{p}_k-\vec{p}_j\right)x\,; \quad x\in[0,1]\,;\quad j,k =1,\,2,\,3\ \land\  j<k\,, 
\label{limit}
\end{equation} 
which correspond to all the possible quantum states with one eigenvalue equal to zero.
The set of states with at least two equal probability values is parametrized by
\begin{equation}
\vec{p}_{jk\ell} = \vec{p}_j + \left(\frac{1}{2}\left[\vec{p}_k+\vec{p}_\ell\right]-\vec{p}_j\right)x\,; \quad x\in[0,1]\,;\quad j\neq k \ \land \ j\neq \ell \ \land \  k\neq\ell\,,
\label{two_equal3}
\end{equation}
and note that $\vec{p}_{jk\ell}=\vec{p}_{j\ell k}$. The centroid of the simplex is given by the point $\vec{p}_e=(1/3,\,1/3,\,1/3)$ representing the most mixed state of the system.

The explicit form of the diagonal density matrix operator in $\lambda$-space is given by
\begin{equation}
\op{\rho} = \diag{\frac{1}{3}+\frac{\lambda_7}{2}+\frac{\lambda_8}{2\sqrt{3}},\frac{1}{3}-\frac{\lambda_7}{2}+\frac{\lambda_8}{2\sqrt{3}},\frac{1}{3}-\frac{\lambda_8}{\sqrt{3}} }.
\end{equation}
which allows us to describe the system by two variables, $\lambda_7$ and $\lambda_8$. For example, the conditions $0 \leq \rho_{jj} \leq 1$ allow us to define the area of the space in which the values of $\lambda_7$ and $\lambda_8$ represent a quantum system. These conditions can be written as
\begin{equation}
-1 \leq \lambda_7 \leq 1, \quad \frac{-2+3\vert \lambda_7 \vert}{\sqrt{3}} \leq \lambda_8 \leq \frac{1}{\sqrt{3}},
\end{equation}
which give a triangle with vertices at  (1, $1/\sqrt{3}$), (-1, $1/\sqrt{3}$), and (0, $-2/\sqrt{3}$), corresponding to the pure states with density matrices $\diag{1,0,0}$, $\diag{0,1,0}$, and $\diag{0,0,1}$, respectively.

Since the simplexes in $p$-space and $\lambda$-space are related through a linear map, we consider first these representations, by choosing the diagonal Gell-Mann operators as
\begin{equation*}
\op{\Lambda}_7 =\diag{1,\,-1,\,0}\,;\qquad
\op{\Lambda}_8 =\frac{1}{\sqrt{3}} \diag{1,\,1,\,-2}\,.
\end{equation*}
Notice that the Gell-Mann vector has been reduced to only two components different from zero. We have the matrix transformation $\vec{p}=\op{M}\vec{\Lambda}$ with $\vec{p}=(p_1,\,p_2,\,p_3)$, $\vec{\Lambda}=(\lambda_7,\,\lambda_8,\,1)$ and 
\begin{equation}\label{eq.matM}
\op{M} = \frac{1}{2}\left(\begin{array}{c c c} 
1& \frac{1}{\sqrt{3}}& \frac{2}{3}\\
-1&\frac{1}{\sqrt{3}} &\frac{2}{3} \\
0& -\frac{2}{\sqrt{3}} &\frac{2}{3}
\end{array}\right)\,, \qquad \op{M}^{-1} = \left(\begin{array}{c c c} 
1& -1& 0\\
\frac{1}{\sqrt{3}}&\frac{1}{\sqrt{3}} &-\frac{2}{\sqrt{3}} \\
1& 1 &1
\end{array}\right)\,,
\end{equation}
and the two-dimensional $\lambda$-space is given by the points $\vec{\lambda}= (\lambda_7,\,\lambda_8)$, so the vertices in the $p-$simplex are mapped to vertices in $\lambda$-space as
\[\vec{p}_1 \to \vec{\Lambda}_1 = \left(1,\frac{1}{\sqrt{3}},\,1\right)\,;\quad \vec{p}_2 \to \vec{\Lambda}_2 = \left(-1,\frac{1}{\sqrt{3}},\,1\right)\,;\quad \vec{p}_3 \to \vec{\Lambda}_3 = \left(0,-\frac{2}{\sqrt{3}},\,1\right)\,;\]
and an arbitrary point by
\[\vec{p}\to \vec{\Lambda}=\left(p_1-p_2,\, \frac{1-3p_3}{\sqrt{3}},\,1\right)\,.\]

Relating again to figure~\ref{f.simplexPL}, we show the values which yield a physical state of the system, both in the simplex Fig.~\ref{f.simplexPL}(a) and in $\lambda$-space Fig.~\ref{f.simplexPL}(b). In both cases the vertices of the triangle correspond to pure states of the system. By considering that the three levels represent an atomic or molecular system and give the level occupation probabilities, we use the following convention: the ground state is indicated by $\vec{p}_1$ or $\vec{\lambda}_1$ (red dot in the figure), the first excited state by $\vec{p}_2$ or $\vec{\lambda}_2$ (plus-symbol in the figure), and the second excited state by $\vec{p}_3$ or $\vec{\lambda}_3$ (times-symbol in the figure). 

Other points of interest are the thermal endpoints corresponding to the most mixed state $\vec{p}_e=(\vec{p}_1+\vec{p}_2+\vec{p}_3)/3$ or $\vec{\lambda}_e=(\vec{\lambda}_1+\vec{\lambda}_2+\vec{\lambda}_3)/3=(0,0)$ (square-symbol in the figure), and the cases when two occupation probabilities of the three-level system are equal (i.e., two eigenvalues of the density matrix are degenerate): $\vec{p}_m=(\vec{p}_i+\vec{p}_j)/2$ or $\vec{\lambda}_m=(\vec{\lambda}_i+\vec{\lambda}_j)/2$, $i\neq j$ (asterisk-symbol in the figure). 

The open (shaded) internal triangle establishes the case without degeneracy, corresponding to the set of points in the simplex for a thermal density matrix; the  black lines connecting the point $\vec{p}_e$ ($\vec{\lambda}_e$) with either the vertex $\vec{p}_1$ ($\vec{\lambda}_1$) or the middle point $\vec{p}_m$ ($\vec{\lambda}_m$) correspond to trajectories from the pure ground state and from the case of a degenerate ground state with two equal occupation probabilities, respectively. The qualitative behavior of the thermal density matrix as a function of the temperature, for the non-degenerate case, is shown by the black curve in the figure, whose concavity increases when the first excited state occupation probability of the system approaches that of the ground state.

\begin{center}
\begin{figure}
\includegraphics[width=0.48\linewidth]{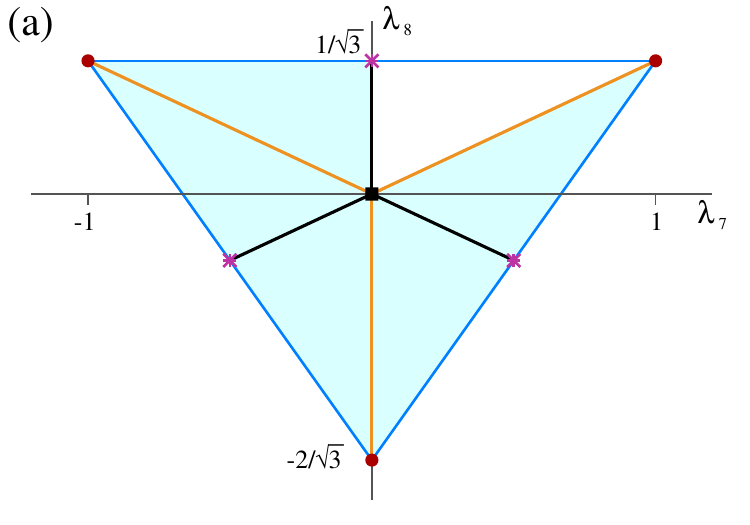}
\includegraphics[width=0.48\linewidth]{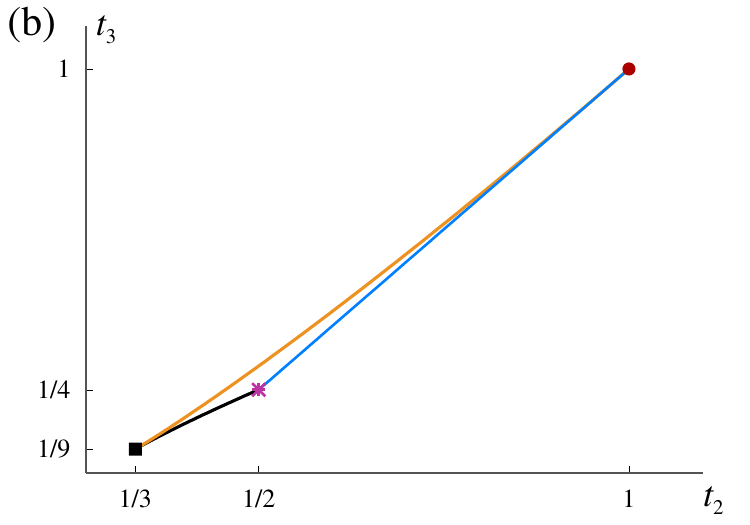}
\caption{Non-linear map from (a) $\lambda$-space to (b) $t$-space. The pure states (dot-symbol) are mapped onto the vertex $\vec{t}_1$ (dot-symbol), the middle-points (asterisk-symbol) onto the vertex $\vec{t}_2$ (asterisk-symbol) and the most mixed state (square-symbol) onto the vertex $\vec{t}_3$ (square-symbol). Each straight-line segment which connects any two indicated points in $\lambda$-space is mapped to a boundary-curve in $t$-space, and each open triangle (as the white triangle) is mapped onto the interior of the bounded region in $t$-space.}
\label{f.mapLT}
\end{figure}
\end{center}

We now consider the non-linear map from $\lambda$-space to $t$-space (cf. Fig.~\ref{f.mapLT}). As pointed out above, this map does not distinguish between pure states, and hence each point in $t$-space represents a set of physical states. In particular, the pure states $\vec{\lambda}_p$ (dot-symbol) at the vertices, middle points $\vec{\lambda}_{m}$ (asterisk-symbol), and the most mixed state $\vec{\lambda}_e$ (square-symbol) indicated in Fig.~\ref{f.mapLT}(a) are mapped to the points  
\[\vec{\lambda}_p\to \vec{t}_1 = (1,1)\,;\quad \vec{\lambda}_m\to \vec{t}_2 = \left(\frac{1}{2},\,\frac{1}{4}\right)\,; \quad \vec{\lambda}_e\to \vec{t}_3 = \left(\frac{1}{3},\,\frac{1}{9}\right)\,, \]
indicated with the same symbols in Fig.~\ref{f.mapLT}(b). Any point $\vec{\lambda}=(\lambda_7,\,\lambda_8)$ in $\lambda$-space is mapped to a point $\vec{t}=(t_2,\,t_3)$ via the transformation
\begin{equation*}
\vec{\lambda}\to \vec{t} = \left(\frac{1}{3} +\frac{1}{2}\left(\lambda_7^2 + \lambda_8^2\right),\, \frac{1}{9} + \frac{\lambda_7^2}{4}\left(2+\sqrt{3}\lambda_8\right)+\frac{\lambda_8^2}{4\sqrt{3}}\left(2\sqrt{3}-\lambda_8\right)\right)\,.
\end{equation*}

The straight line segments connecting points in $\lambda$-space, shown in Fig.~\ref{f.mapLT}(a), can be parametrized as follows:
\begin{eqnarray*}
\vec{\lambda}^{ep}_{j} &=& \vec{\lambda}_j\,x\,;\qquad \vec{\lambda}^{em}_{jk} = \frac{1}{2}\left(\vec{\lambda}_j+\vec{\lambda}_k\right)\,x\,; \quad x\in[0,1] \,,\\[2mm]   
\vec{\lambda}_{jk}^{mp} &=& \vec{\lambda}_j +  \left(\vec{\lambda}_k-\vec{\lambda}_j\right)\,x\,; \quad x\in\left[\frac{1}{2},1\right]\,,
\end{eqnarray*}
for $j,k=1,2,3$, with the upper-indices indicating the corresponding connected points: most mixed state ($e$), middle point of two occupation probabilities ($m$), and pure states $(p)$. Each path maps onto a trajectory in $t$-space corresponding to their boundaries. Any point in $t$-space lying within these trajectories is physically aceptable (and only those), since the density matrix is given in a diagonal form. After the transformation these segments will be given by [see Fig.~\ref{f.mapLT}(b)]
\begin{eqnarray*}
\vec{t}_1^{\ ep} &=&\frac{1}{9}\left(3 + 6 x^2,\, 1+6x^2+2 x^3\right)\,;\quad x\in[0,1]\,,\\[2mm]
\vec{t}_{12}^{\ em}&=& \frac{1}{3}\left(1+\frac{x^2}{2},\, \frac{1}{3} + \frac{x^2}{2} - \frac{x^3}{12}\right)\,;\quad x\in[0,1]\,,\\[2mm]
\vec{t}_{21}^{\ mp} &=&\left(1-2x+2x^2,\, 1-3x+x^2\right)\,;\quad x\in\left[\frac{1}{2},1\right]\,.
\end{eqnarray*}
Notice that the open internal triangle in the $\lambda$-space [white triangle in Figure~\ref{f.mapLT}(a)], representing the domain of thermal states as was discussed previously [cf. Figs.~\ref{f.simplexPL}(b) and \ref{f.mapLT}(a)], is mapped onto the full open region in $t$-space bounded by the trajectories shown. In fact, all points $\vec{t}$ with the exception of the boundary that connects the vertices $\vec{t}_1$ and $\vec{t}_2$ may be associated to a thermal state. The same happens for all other triangles in $\lambda$-space.

The non-linear map from the Gell-Mann variables to the invariants representation has been given above; however, it is interesting to determine the set of density matrices with a selected constant value for an invariant, i.e., the set with a fixed value for $t_2 = {\rm Tr}\,[\op{\rho}^2]$ (purity), or fixed value for $t_3={\rm Tr}\,[\op{\rho}^3]$. The first condition  $p_1^2+p_2^2+p_3^2=t_2$ gives a circumference centered at $\vec{p}_e$ on the simplex, which can be parametrized by
\begin{equation}\label{eq.t2c}
\vec{p}_{t_2} = \vec{p}_e + \sqrt{\frac{3t_2-1}{3}}\left[\cos(\alpha)\,\hat{e}_1 + \sin(\alpha)\,\hat{e}_2\right]\,; \quad \alpha\in\left[0,2\pi\right]\,,
\end{equation}
where the orthonormal vectors $\hat{e}_1$ and $\hat{e}_2$ may be obtained from the diagonal Gell-Mann matrices (see~\ref{ap.pvectors}), and restricts the values for $t_2$ to $1/3\leq t_2\leq 1$\,.

The case $t_3\,=\,$constant is given by the set of points of the form
\begin{equation}\label{eq.t3c}
\vec{p}_{t_3} = \vec{p}_e + r(t_3,\alpha)\left[\cos(\alpha)\,\hat{e}_1 + \sin(\alpha)\,\hat{e}_2\right]\,; \quad \alpha\in\left[0,2\pi\right]\,,
\end{equation}
where $r(t_3,\alpha)$ satisfies
\[\frac{1}{9}+r^2 + \frac{\cos(3\alpha)}{\sqrt{6}}r^3 = t_3\,,\]
and the constraints for having physical states are
\[\frac{1}{9}\leq t_3\leq 1\,,\qquad 0\leq r\leq \sqrt{\frac{2}{3}}\,.\]
The latter are obtained by using the boundaries of the invariant $t_2$ and the fact that $r$ is the radius in a polar coordinate vector representation (in this case is proportional to the length of Bloch vector), so that $\vec{p}_{t_3}$ is a single-valued vector function of $\alpha$. Additional to these constraints, we know that only points on the simplex represent physical states since $p_1+p_2+p_3=1$. 

In $\lambda$-space the same set of trajectories are obtained from the linear map $\vec{\Lambda}=\op{M}^{-1}\vec{p}\,$ with $\op{M}$ given in Eq.~(\ref{eq.matM}). So the parametric curves in $\lambda$-space read $\vec{\Lambda}_{jk} = \op{M}^{-1}\vec{p}_{jk}$, $\vec{\Lambda}_{jk\ell} = \op{M}^{-1}\vec{p}_{jk\ell}$, $\vec{\Lambda}_{t_2} = \op{M}^{-1}\vec{p}_{t_2}$ and $\vec{\Lambda}_{t_3} = \op{M}^{-1}\vec{p}_{t_3}$, respectively.

\begin{figure}
\centering
\includegraphics[width=0.48\linewidth]{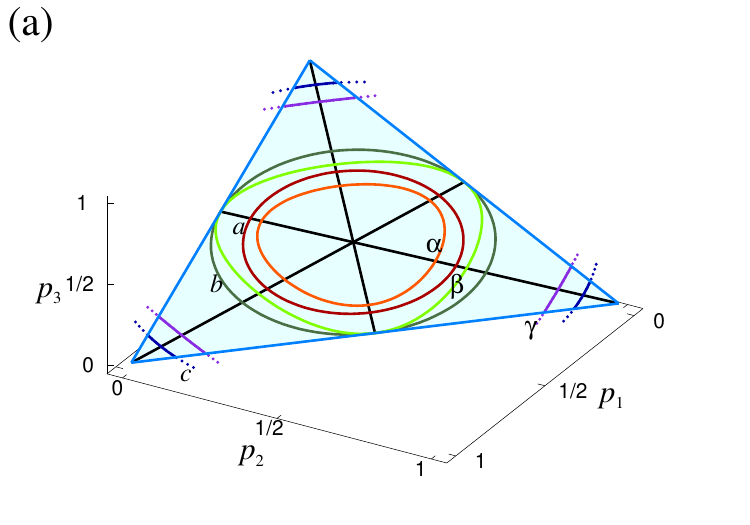}
\includegraphics[width=0.48\linewidth]{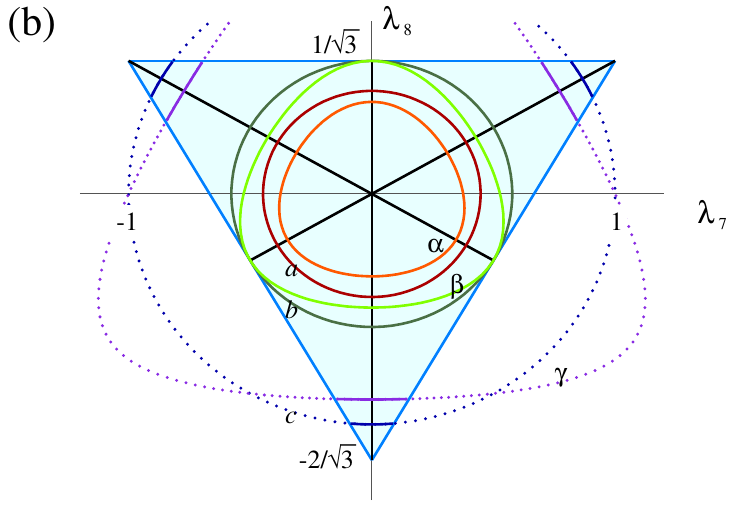}
\caption{Simplex in $p$-space (a), and affine $\lambda$-space (b). The edges of the simplex given by Eq.~(\ref{limit}) correspond to states with at least one eigenvalue equal to zero; the medians of the triangle given by Eq.~(\ref{two_equal3}) contain states with at least two degenerate eigenvalues; the centered circles given by Eq.~(\ref{eq.t2c}) represent the set of density matrices with equal purity (invariant $t_2$), shown here for $t_2=13/30,1/2,5/6$ (labeled by $a,\,b$ and $c$, respectively); while the triangles with rounded corners given by Eq.~(\ref{eq.t3c}) represent the set of density matrices with fixed invariant $t_3$, here shown for $t_3=109/600,1/4,11/18$ (labeled by $\alpha,\,\beta$ and $\gamma$, respectively). The corresponding ones, via the linear transformation, are shown at right in $\lambda$-space.}
\label{fig1}
\end{figure}

Figure \ref{fig1} shows all the properties mentioned above for the qutrit system in $p$-space [Fig.~\ref{fig1}(a)] and in $\lambda$-space [Fig.~\ref{fig1}(b)].
The edges of the simplex given by Eq.~(\ref{limit}) represent the set of density matrices with at least one of their eigenvalues equal to zero; the medians of the triangle (black lines) parametrized by Eq.~(\ref{two_equal3}) correspond to the set of density matrices with at least two degenerate eigenvalues; the curves (or segments of curve thereof) inside the simplex represent the set of states with one constant invariant. For $t_2$, corresponding to circles with center at the most mixed state Eq.~(\ref{eq.t2c}), we have labeled a few cases:  $(a)$ for $t_2=13/30$, $(b)$ for $t_2=1/2$, and $(c)$ for $t_2=5/6$. For $t_3$ constant Eq.~(\ref{eq.t3c}) we have trajectories that look like triangles with rounded corners (resembling a Reuleaux triangle) centered at the most mixed state, and we have labeled them ($\alpha$) for $t_3=109/600$, ($\beta$) for $t_3=1/4$, and ($\gamma$) for $t_3=11/18$. In both cases we may see the continuation outside the simplex (unphysical states) of the broken trajectories, drawn here with dots in order to illustrate the shape of the full curve in the whole space.

\begin{figure}
\begin{center}
\includegraphics[width=0.55\linewidth]{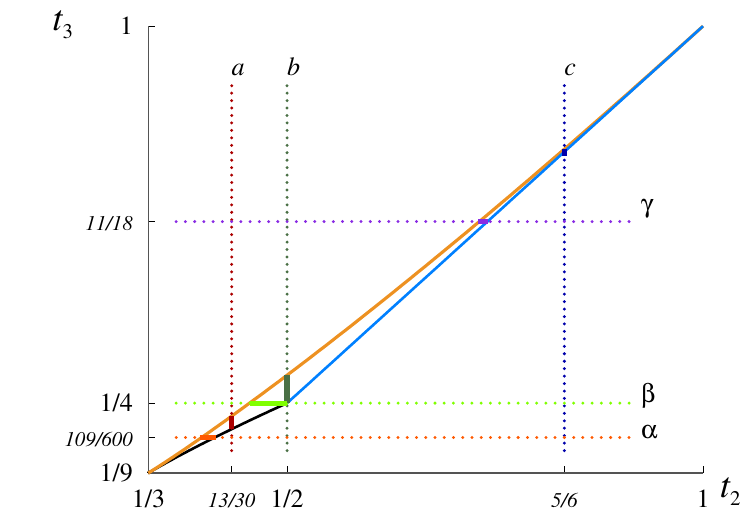}
\end{center}
\caption{Physical region in $t$-space obtained via the non-linear map from the simplex in $p$-space or the affine domain in $\lambda$-space [Eqs.~(\ref{limit})--(\ref{eq.t3c})]. The image of the curves in Figure~\ref{fig1} carry the same labels here. The unphysical states along the constant dotted lines are drawn for the purpose of clarity. The boundary of the region is given by $t_3=t_2-\frac{2}{9}\pm\frac{(3t_2-1)^{3/2}}{9\sqrt{2}}$, (upper (orange) and short lower (black) lines respectively) and are associated to density matrices with two equal eigenvalues, while the long lower (blue) curve is described by $t_3=\frac{1}{2}(3t_2-1)$ and is associated to density matrices with an eigenvalue equal to zero.}
\label{fig2}
\end{figure}

Now for the non-linear map from $\lambda$-space to $t$-space. The centered circumferences in $\lambda$-space or $p$-space map onto vertical cutoff lines in $t$-space, while the kind of trajectories given by Eq.~(\ref{eq.t3c}), which resemble a kind of Reuleaux triangle, map onto horizontal cutoff lines in $t$-space. These are shown in Figure~\ref{fig2}, where we have labeled the trajectories in a similar fashion to Figure~\ref{fig1}. For clarity, we have extended the trajectories out of the physical region with dotted lines; the counterpart to these extensions, which are unphysical, do not appear in the non-linear map of a single inscribed curve in the affine domain of $\lambda$-space, since any inscribed curve in the affine domain maps to a curve in the physical $t$-space. The boundary of the compact region in Figure~\ref{fig2} corresponds to the edges and medians of the triangle (see Fig.~\ref{f.mapLT}), i.e., they are composed by states with at least one vanishing eigenvalue (vertices and edges of the triangle in $p$- or $\lambda$-space) and states with degenerate eigenvalues (medians of the triangle in $p$- or $\lambda$-space).   

Summarizing, we have represented a qutrit by a $3 \times 3$ matrix yielding three eigenvalues in its diagonal representation, i.e., $\op{\rho}_D=\diag{p_1,p_2,p_3}$. The diagonal state can be studied in several ways, by examining directly the three eigenvalues ($p$-space), equivalently in a $2$-dimensional $\lambda$-space by using the Gell-Mann parametrization, or
 by exploring the invariants of the density matrix $t_2={\rm Tr} \, \op{\rho}_D^2$ and $t_3={\rm Tr} \, \op{\rho}_D^3$ ($t$-space).

\subsection{Ququart system}

\begin{figure*}
\begin{center}
\includegraphics[width=0.48\linewidth]{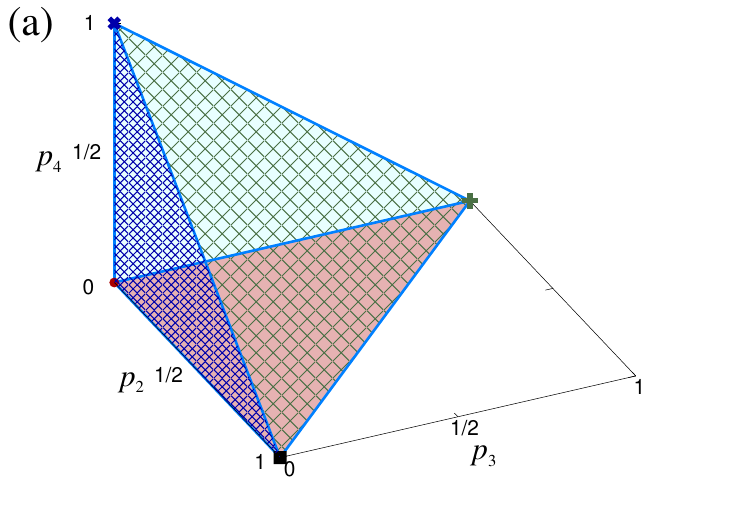}
\includegraphics[width=0.48\linewidth]{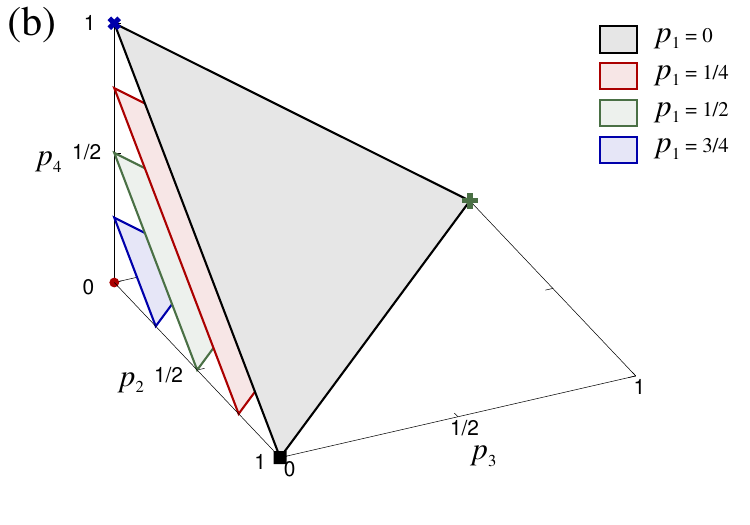}\\
\includegraphics[width=0.48\linewidth]{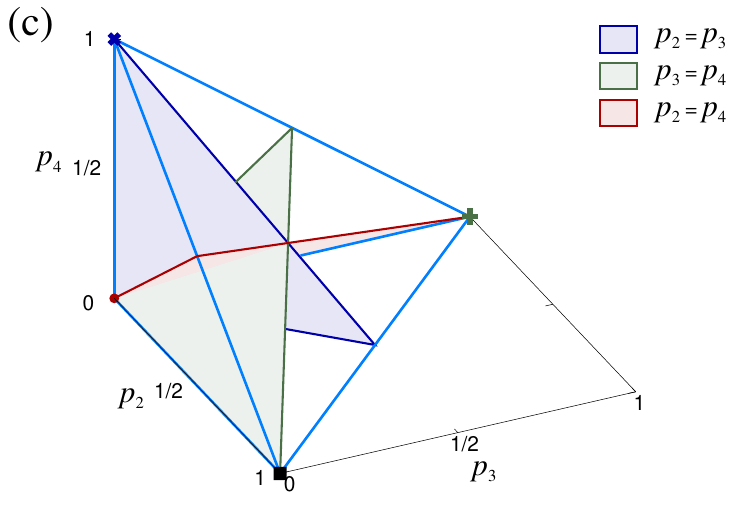}
\caption{Simplex for a ququart system in $p$-space. (a) Solid region in the subspace $\{p_2,\,p_3,\,p_4\}$; (b) the simplex for fixed values of probability $p_1$; and (c) the cut planes when two probabilities take the same value. In all cases the pure states are indicated by dot-symbols for $\op{\rho}_1$, square-symbols for $\op{\rho}_2$, plus-symbols for $\op{\rho}_3$ and times-symbols for $\op{\rho}_4$.}
\label{f.simplex2P}
\end{center}
\end{figure*}

The diagonal density matrix of a ququart system is given in $p$-space as Eq.~(\ref{eq.rhoP}) 
\[\op{\rho}=\diag{p_1,\,p_2,\,p_3,\,p_4}=\diag{\vec{p}}\,,\]
its $3$-simplex in ${\rm U}(4)$ is a tetrahedron with vertices at $\op{\rho}_j=\diag{\vec{p}_j}$ for which $p_j=1$. It can be visualized as a solid in $3$-dimensional space, generated by the probabilities $p_2,\,p_3$ and $p_4$, with $p_1=1-p_2-p_3-p_4$ as in figure~\ref{f.simplex2P}(a). Its vertices, indicated by dot-, square-, plus- and times-symbols, correspond to pure states; according to $\op{\rho}_j=\diag{\vec{p}_j}$ these are denoted by $\op{\rho}_1,\, \op{\rho}_2,\,\op{\rho}_3$ and $\op{\rho}_4$. The faces correspond to the set of points where at least one eigenvalue is zero, while at the edges at least two eigenvalues vanish. Therefore, one has double degeneracy corresponding (as a limit case, see discussion below) to the orbits in ${\rm U}(4)$ of dimension 6, associated to the quotient group ${\rm U}(4)/\left({\rm U}(2)\otimes {\rm U}(1)\otimes {\rm U}(1)\right)$.
For a fixed value of $p_1$ the set given by
\[p_2+p_3+p_4=c\geq 0\,;\qquad c:=1-p_1\,,\]
defines a region similar to that in the case of the qutrit [cf. Fig.~\ref{f.simplexPL}(a)] which grows as $p_1 \to 0$ (cf. Fig.~\ref{f.simplex2P}(b)). The case when two probability values are equal, given by the condition
\[p_j+p_k+2p_\ell =1\,, \qquad p_m=p_\ell\,,\]
cuts the simplex at a plane; the cases $p_m=p_\ell$ with $m\neq 1$ and $\ell\neq 1$ are shown in figure~\ref{f.simplex2P}(c). These planes in turn generate the medians of the triangular face for a constant value $p_1$. The most mixed state lies at
\[\vec{p}_e:=\frac{1}{4}\left(1,\,1,\,1,\,1\right)\,.\]

One may then parametrize each point of the simplex in the $p$-space as 
\begin{equation}\label{eq.vpr}
\vec{p}= \vec{p}_e + \frac{r}{\sqrt{2}} \left[\cos(\phi)\sin(\theta)\,\hat{e}_1 + \sin(\phi)\sin(\theta)\,\hat{e}_2+\cos(\theta)\,\hat{e}_3\right]\,,
\end{equation}
with $r\geq0, \phi\in[0,\,2\pi]$ and $\theta\in[0,\,\pi]$ and where the orthonormal vectors were obtained from the diagonal Gell-Mann matrices (see~\ref{ap.pvectors}).

\begin{figure}
\begin{center}
\includegraphics[width=0.48\linewidth]{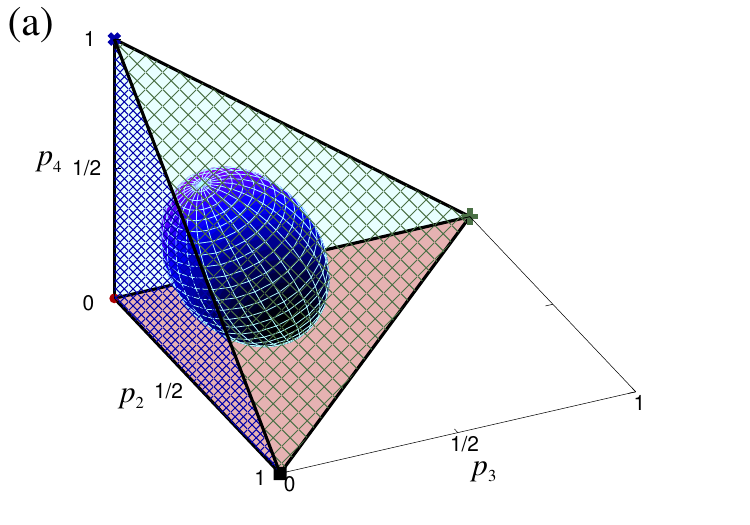}
\includegraphics[width=0.48\linewidth]{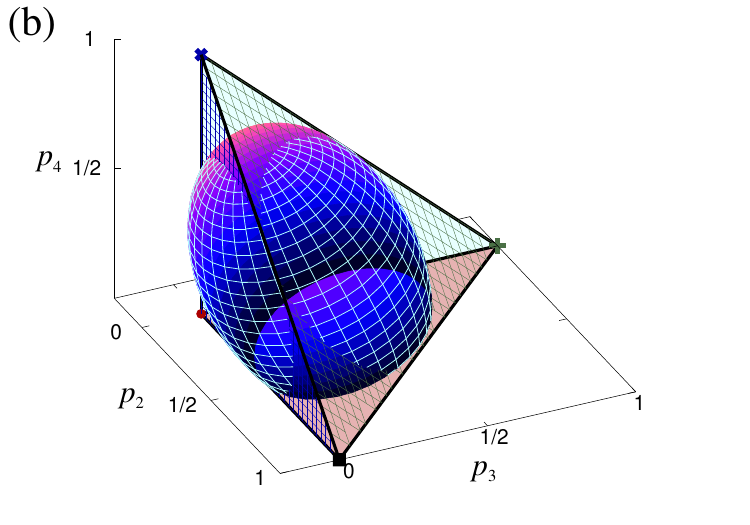}\\
\includegraphics[width=0.48\linewidth]{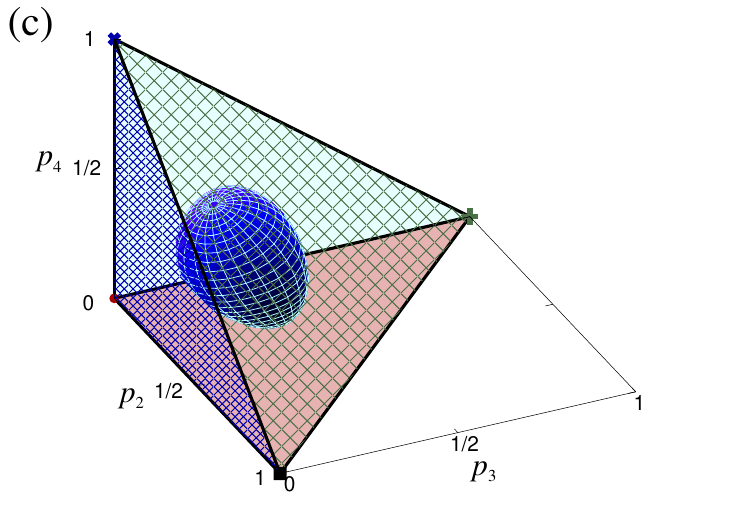}
\includegraphics[width=0.48\linewidth]{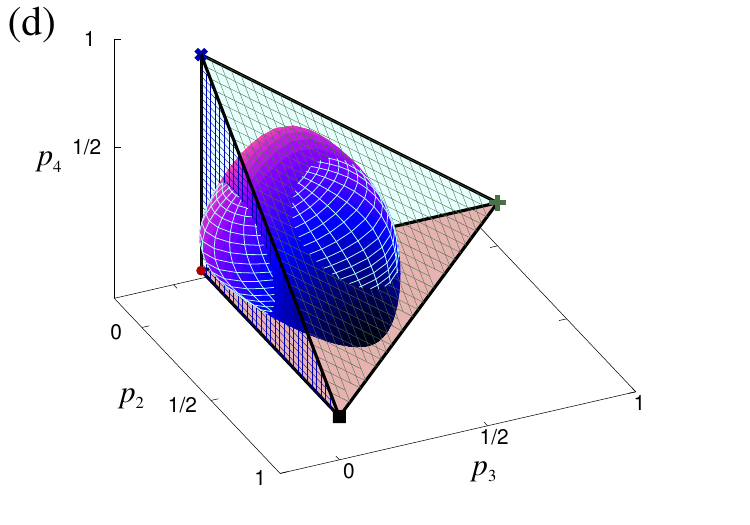}\\
\includegraphics[width=0.48\linewidth]{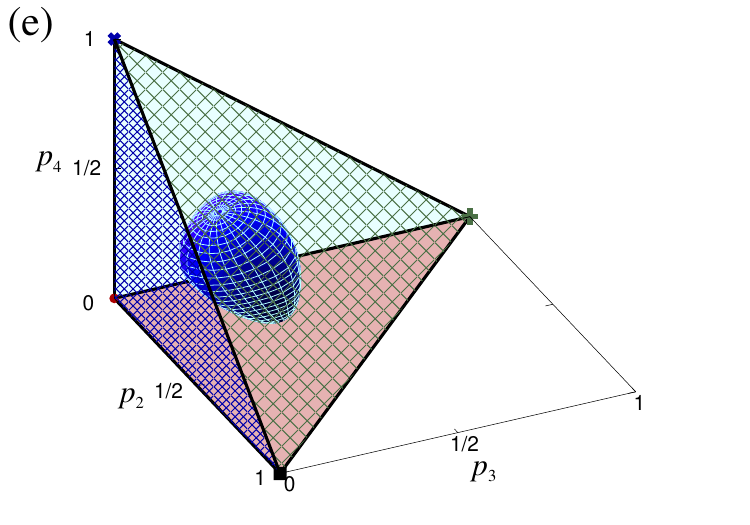}
\includegraphics[width=0.48\linewidth]{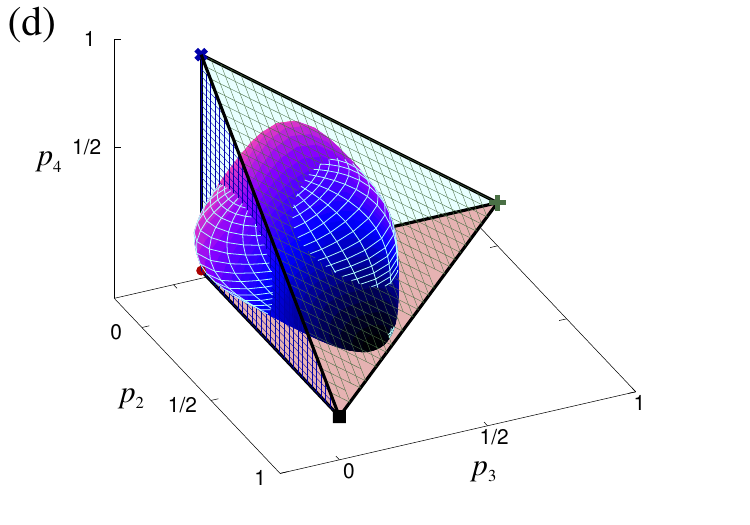}
\end{center}
\caption{Surfaces defined on the $3-$simplex for the ququart when one of the invariants of ${\rm SU}(4)$ takes a constant value: (a) $t_2=1/3$, (b) $t_2=1/2$, (c) $t_3=1/10$, (d) $t_3=7/40$ (e) $t_4=1/32$ and (f) $t_4=5/64$. Points outside the simplex correspond to unphysical states.}
\label{f.simplex3P}
\end{figure}

The advantage of writing the point in the simplex in this form is to obtain the invariants in terms of the length of the Bloch vector $r$ and polar variables, so a density matrix which is written in a general form as $\op{\rho}=\diag{\vec{p}}$ has the invariants
\begin{eqnarray}
t_2 &=& \frac{1}{4}\left(1 + 2r^2\right)\,;\qquad t_3 = \frac{1}{16} + \frac{3}{8} r^2 + \frac{a_3}{96}  r^3\,; \nonumber\\[2mm]
t_4 &=& \frac{1}{64}+\frac{3}{16}r^2 + \frac{a_3}{96}r^3 + \frac{b_4}{384}r^4\,, \label{eq.maptQq}
\label{p_to_t_b}
\end{eqnarray}
where we have used
\begin{eqnarray*}
a_3 &:=&-\sqrt{6}\left[3\cos(\theta)+5\cos(3\theta)\right]+8\sqrt{3}\sin^3(\theta)\sin(3\phi)\,;\\
b_4 &:=&45+4\cos(2\theta)+7\cos(4\theta)+32\sqrt{2}\cos(\theta)\sin^3(\theta)\sin(3\phi)\,,
\end{eqnarray*}
to simplify the notation. Thus, in $p$-space the condition $t_2$ constant yields a hyper-sphere of radius $r=\sqrt{4t_2-1}/2$ centred at $\vec{p}_e$. On the simplex, represented by the coordinate axes $p_2,\,p_3$ and $p_4$, this hyper-sphere is seen elongated: at the limit values of $t_2$ one obtains the point $\vec{p}_e$ ($t_2=1/4$) and a surface which intersects the simplex only at the vertices ($t_2=1$); intermediate values generate a surface lying completely inside the simplex. Figure~\ref{f.simplex3P} shows this for different values of $t_2$: (a) $t_2=1/3$; (b) $t_2=1/2$. Points outside the simplex correspond to unphysical states.

Solving for $t_3$ and $t_4$ in terms of $r\geq0$ one finds the parametric form of the surfaces of constant value for each of the invariants. In both cases one has a symmetry of $2\pi/3$ in the angular variable $\phi$ as expected by the form of the coefficients $a_3$ and $b_4$. These surfaces are shown in Figure~\ref{f.simplex3P}, (c) for $t_3=1/10$, (d) for $t_3=7/40$, (e) for $t_4=1/32$ and (f) for $t_4=5/64$.

The explicit density matrix in $\lambda$-space can be written as
\begin{equation}
\fl \op{\rho} = \diag{\frac{1}{4}+\frac{\lambda_{13}}{2}+\frac{\lambda_{14}}{2\sqrt{3}}+\frac{\lambda_{15}}{2\sqrt{6}},\ \frac{1}{4}-\frac{\lambda_{13}}{2}+\frac{\lambda_{14}}{2\sqrt{3}}+\frac{\lambda_{15}}{2\sqrt{6}},\ \frac{1}{4}-\frac{\lambda_{14}}{\sqrt{3}}+\frac{\lambda_{15}}{2\sqrt{6}},\ \frac{1}{4}-\frac{3\lambda_{15}}{2\sqrt{6}} }\,.
\end{equation}
Since the density matrix is positive semi-definite $0\leq \rho_{jj} \leq 1$ ($j=1,\ldots,4$), we have a tetrahedron given by
\begin{equation}
-1 \leq \lambda_{13} \leq 1, \quad -\frac{2}{\sqrt{3}} \leq \lambda_{14} \leq \frac{1}{\sqrt{3}}, \quad -\sqrt{\frac{2}{3}} \leq \lambda_{15} \leq \frac{1}{\sqrt{6}}.
\end{equation}
The pure states, corresponding to the density matrices $\op{\rho} = {\rm diag}(1,0,0,0)$, $\op{\rho} = {\rm diag}(0,1,0,0)$, $\op{\rho} = {\rm diag}(0,0,1,0)$, and $\op{\rho} = {\rm diag}(0,0,0,1)$, are associated to the vertices $\left(1,\frac{1}{\sqrt{3}},\frac{1}{\sqrt{6}}\right)$, $\left(-1,\frac{1}{\sqrt{3}},\frac{1}{\sqrt{6}} \right)$, $\left(0, -\frac{2}{\sqrt{3}},\frac{1}{\sqrt{6}} \right)$, and $\left(0,0,-\sqrt{\frac{2}{3}} \right)$, respectively. The values for $\lambda_{13}$, $\lambda_{14}$, $\lambda_{15}$ on the faces of the tetrahedron correspond to pure states, while the centroid corresponds to the most mixed state with $\op{\rho} ={\rm diag}(1/4,1/4,1/4,1/4)$.

To visualise the thermal region it is convenient to consider the Gell-Mann representation, by choosing the diagonal matrices as 
\begin{eqnarray*}
\op{\Lambda}_{13} &=&\diag{1,\,-1,\,0,\,0}\,;\qquad
\op{\Lambda}_{14} = \frac{1}{\sqrt{3}}\diag{1,\,1,\,-2,\,0}\,;\\[2mm]
\op{\Lambda}_{15} &=&\frac{1}{\sqrt{6}} \diag{1,\,1,\,1,\,-3}\,;
\end{eqnarray*}
each point $\vec{\Lambda}:=(\lambda_{13},\,\lambda_{14},\,\lambda_{15},\,1)$ is related to the vector $\vec{p}=(p_1,\,p_2,\,p_3,\,p_4)$ in probability space via the linear map
\begin{equation}
\vec{p}=\op{M}\vec{\Lambda}\,; \qquad \vec{\Lambda}=\op{M}^{-1}\vec{p}
\end{equation}
where the matrix transformation is
\begin{equation}
\op{M} = \frac{1}{2}\left(
\begin{array}{c c c c}
1 & \frac{1}{\sqrt{3}}&  \frac{1}{\sqrt{6}}& \frac{1}{2} \\[2mm]
-1& \frac{1}{\sqrt{3}} & \frac{1}{\sqrt{6}}& \frac{1}{2}\\[2mm]
0 & - \frac{2}{\sqrt{3}} & \frac{1}{\sqrt{6}}& \frac{1}{2}\\[2mm]
0 & 0 &-\sqrt{\frac{3}{2}} & \frac{1}{2}
\end{array}
\right)\,;\quad 
\op{M}^{-1} = \left(
\begin{array}{c c c c}
 1& -1&0 & 0\\[2mm]
 \frac{1}{\sqrt{3}} & \frac{1}{\sqrt{3}} &- \frac{2}{\sqrt{3}} &0 \\[2mm]
\frac{1}{\sqrt{6}} & \frac{1}{\sqrt{6}}& \frac{1}{\sqrt{6}}&-\sqrt{\frac{3}{2}} \\[2mm]
1 &1 &1 &1 
\end{array}
\right)\,,
\end{equation}
and the $\lambda$-space is defined by the points of the form $\vec{\lambda}=(\lambda_{13},\,\lambda_{14},\,\lambda_{15})$.

\begin{figure}
\begin{center}
\includegraphics[width=0.48\linewidth]{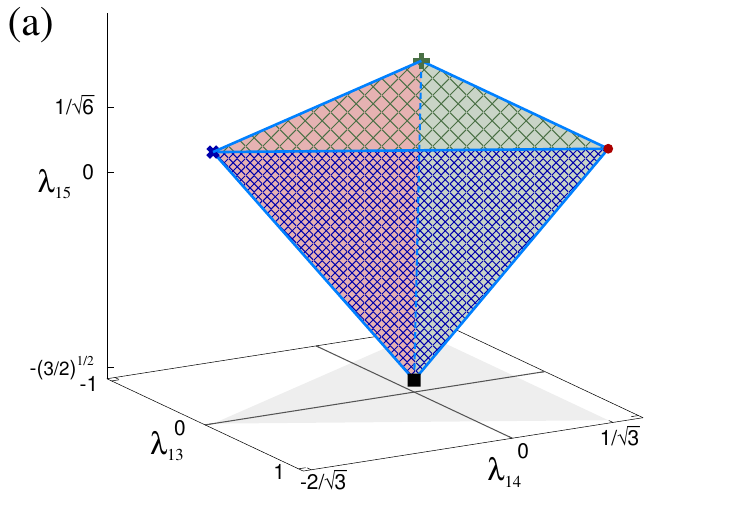}
\includegraphics[width=0.48\linewidth]{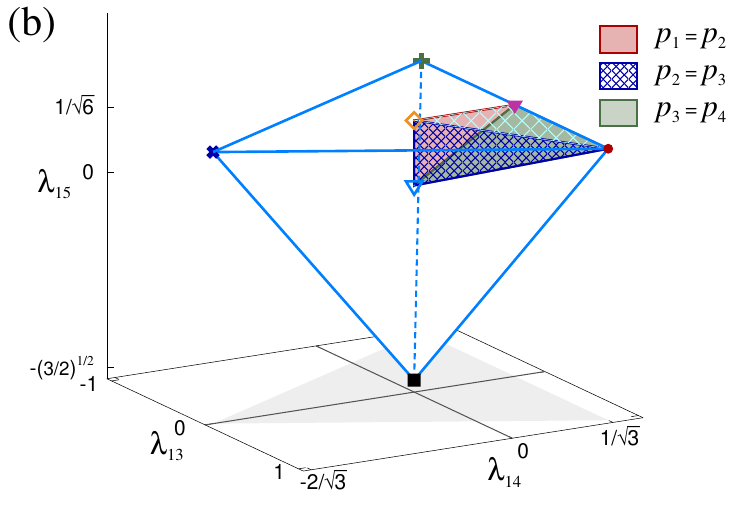}\\
\includegraphics[width=0.48\linewidth]{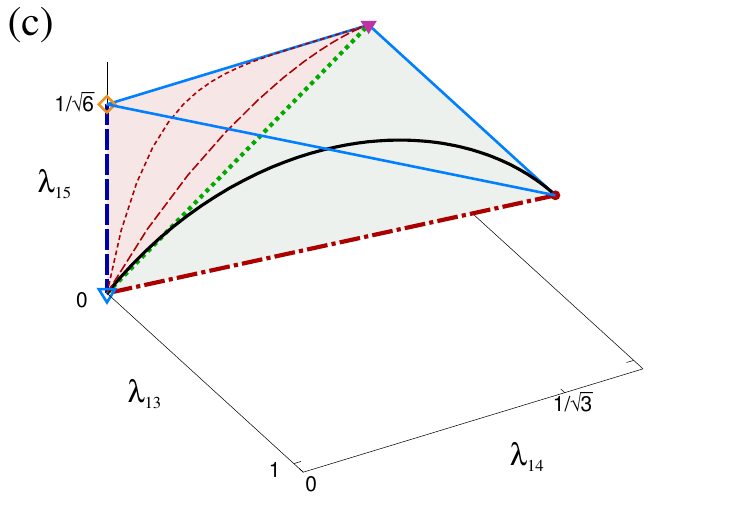}
\includegraphics[width=0.48\linewidth]{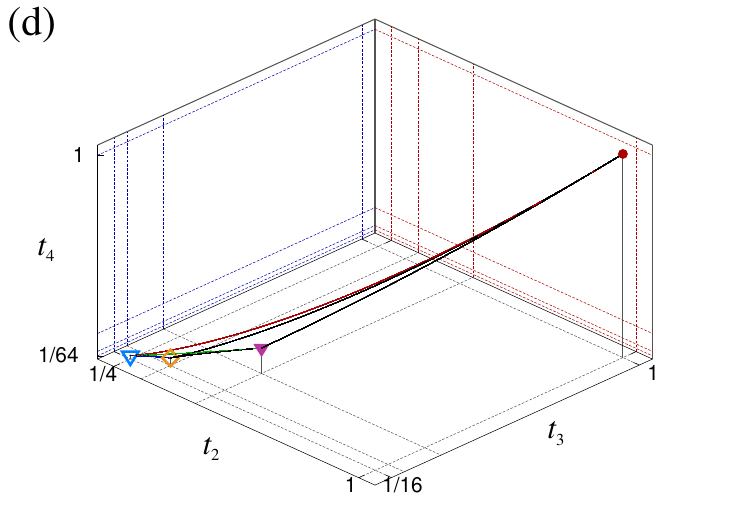}
\end{center}
\caption{(a) Simplex for a ququart system in $\lambda$-space. (b) Thermal state region by choosing the vertices to represent probabilities for ground state (dot-symbol), first excited state (plus-symbol), second excited state (times-symbol) and third excited state (square-symbol). Other points indicated are for the most mixed state (empty triangle-symbol), the end point for degeneracy $p_1=p_2$ (solid triangle-symbol) and the end point for degeneracy $p_1=p_2=p_3$ (empty square-symbol). (c) Zoom-in of the thermal states region with some trajectories as functions of the temperature. (d) The physical states region in $t$-space obtained from the non-linear map applied to the thermal region in $\lambda$-space. See text for details.}
\label{f.simplex3L} 
\end{figure}

From Eq.(\ref{eq.vpr}) where the unit vectors $\hat{e}_j$ are defined to parametrize the probability vector $\vec{p}$, and from the relation between the latter and the diagonal Gell-Mann matrices $\op{\Lambda}_{jk}$, it is clear that any hypersurface $t_2$ constant is mapped onto a sphere in $\lambda$-space. The affine space of states in $\lambda$-space is a tetrahedron which is shown in Figure~\ref{f.simplex3L}(a), where the vertices correspond to pure states indicated by dot-, plus-, times-, and square-symbols respectively. Under the map these are
\begin{eqnarray*}
\vec{p}_1\to \vec{\lambda}_1 &=&\left(1,\,\frac{1}{\sqrt{3}},\,\frac{1}{\sqrt{6}}\right)\,;\quad \vec{p}_2\to \vec{\lambda}_2 =\left(-1,\,\frac{1}{\sqrt{3}},\,\frac{1}{\sqrt{6}}\right)\,;\\[2mm]
\vec{p}_3\to \vec{\lambda}_3 &=&\left(0,\,-\frac{2}{\sqrt{3}},\,\frac{1}{\sqrt{6}}\right)\,;\qquad \vec{p}_4\to \vec{\lambda}_4 =\left(0,\,0,\,-\sqrt{\frac{3}{2}}\right)\,,
\end{eqnarray*}
which yield the limit values of $\vec{\lambda}$ for physical states.
Referring to this figure, we may visualize the thermal states by assigning  a definite meaning to each vertex; thus, we choose $p_1$ to denote the probability of the ground state (dot-symbol), $p_2$ for the probability of the first excited state (plus-symbol), $p_3$ for that of the second excited state (times-symbol), and $p_4$ for that of the third excited state (square-symbol). This would give the condition $p_1\geq p_2\geq p_3\geq p_4$. The planes defined by $p_1=p_2$, $p_2=p_3$ and $p_3=p_4$ are mapped onto planes in $\lambda$-space and the intersection with the affine space of states provides the faces of the thermal region shown in Fig.~\ref{f.simplex3L}(b). In this sense, with the exception of the upper face, each point in this region is related to a density matrix of the form Eq.~(\ref{therm_gen}). As mentioned before, for infinite temperature the thermal state corresponds to the most mixed state, indicated by an empty triangle-symbol, and for zero temperature the thermal state corresponds to a ground state (dot-symbol) without degeneracy; the case for the ground state with degeneracy is indicated by a triangle-symbol (degeneracy of order two) and empty square-symbol (degeneracy of order three), respectively. So, the thermal state as a function of the temperature moves along a trajectory which connects the most mixed state with one of the other three end points depending on the degeneracy of the system.
Fig.~\ref{f.simplex3L}(c) shows these paths: a solid black line for the general case without degeneracy, while for the degenerate cases the path lies on a face or edge of the affine space; for example, for values $p_1=p_2$ with $p_3\neq p_1$ and $p_3\neq p_4$ the trajectory lies on the face $\lambda_{13}=0$ (drawn with a dashed-line for $p_3\approx p_4$, and with a short-dashed-line for $p_3\approx p_1$). A similar behavior is observed on the other thermal faces defined by the degeneracy $p_2=p_3$ and $p_3=p_4$, with the trajectories on the edges corresponding to the cases for double degeneracy $p_1=p_2$ and $p_3=p_4$ (dotted-line), degeneracy of order three $p_1=p_2=p_3$ (dashed-line), and when $p_2=p_3=p_4$ (dash-dot-line). 

Finally, for $\vec{\lambda} = r\left(\cos(\phi)\sin(\theta),\, \sin(\phi)\sin(\theta),\,\cos(\theta)\right)$ in spherical coordinates, the non-linear map from $\lambda$-space to the invariants $t$-space $\vec{\lambda}\to\vec{t}=(t_2,\,t_3,\,t_4)$ gives us the values $t_j$ shown by Eq.~(\ref{eq.maptQq}). The end points of the thermal region are mapped to the vertices of the physical region in $t$-space [cf. Eq.~(\ref{eq.vertices.t})], i.e., these points are representative of a particular kind of states in $t$-space. In $\lambda$-space these points are given by $\vec{\lambda}_p=(1,\,1/\sqrt{3},\,1/\sqrt{6})$ (ground state),  $\vec{\lambda}_{12} =(0,\,1/\sqrt{3},\,1/\sqrt{6})$ (the degenerate case $p_1=p_2$), $\vec{\lambda}_{123} =(0,\,0,\,1/\sqrt{6})$ (the degenerate case $p_1=p_2=p_3$) and $\vec{\lambda}_e =(0,\,0,\,0)$ (the most mixed state), and these are mapped to $\vec{\lambda}_p\to\vec{t}_1$, $\vec{\lambda}_{12}\to\vec{t}_2$, $\vec{\lambda}_{123}\to\vec{t}_3$ and $\vec{\lambda}_e\to\vec{t}_4$; these points  together with the boundary of the thermal state region are shown in Fig.~\ref{f.simplex3L}(d).

\section{Applications to Lipkin-type Hamiltonian models}
\label{s-thermal}

The different diagonal representations discussed above allow us to study several processes related to three-- and four--level systems. For example, one may study the thermodynamics of the Gibbs states represented by Eq.~(\ref{therm_gen}) for different Hamiltonians. In this section, we present the study for two cases: where the Hamiltonian is given as a linear form of the angular momentum generators $\op{J}_x $, $\op{J}_y$, and $\op{J}_z$; and in the quadratic form described by the LMG Hamiltonian model.

\subsection{Linear Hamiltonian}

The linear Hamiltonian can be written as
\begin{equation}
H = \omega \ \hat{n} \cdot \op{J} \, ,
\label{eq.Hdiag}
\end{equation}
where $\hat{n}={(\sin\theta \cos\phi, \sin\theta \sin\phi, \cos\theta)}$ denotes the unit vector in spherical coordinates. By means of the basis states of $\op{J}^2$ and $\op{J}_n$ denoted by $\vert J, M\rangle_n$ the equidistant eigenvalues and eigenvectors are given by the expressions ($\hbar=1$)
\begin{equation}
E_M=\omega \,  M  \ ;\quad | J, M \rangle_n = \sum^J_{\mu=-J}  D^J_{\mu, M}(0,-\theta,\phi) | J,\, \mu \rangle \, ,  \quad M=-J,-J+1,\cdots J-1,J\, ,
\label{eq.EM}
\end{equation}
where $D^J_{\mu,M}$ denotes the elements of the Wigner rotational matrix. 

For this system, the PME of an ensemble of particles of spin $\op{J}$ and with a well defined mean value of the energy yields a density matrix of the form (cf. Section~\ref{s.VM})
\begin{equation}
\rho= \diagb{p_1,p_2, \cdots,p_{2 J}, p_{2 J+1}} \, ,
\label{eq1a}
\end{equation}
with occupation probability for a thermal state
\begin{equation}
p_k = \frac{e^{-\beta \, E_{k -J-1}}}{\sum^J_{M=-J} e^{-\beta \, E_M}} = \frac{1}{Z}e^{-\beta \, E_{k -J-1}}\,,\quad k=1,\,2,\,\dots\,,2J+1\,,
\label{eq1b}
\end{equation}
with $Z$ the partition function. Since the spectrum is equidistant for this choice of labels one has the relation
\begin{equation}\label{eq.pkpK}
p_{k}\,p_{2(J+1)-k} = \frac{1}{Z^2}\,.
\end{equation}

The expressions~(\ref{eq1a}) and (\ref{eq1b}) are used to describe the thermal states. Particular cases with $J=1$ for a qutrit and $J=3/2$ for a ququart will be considered.

\subsection*{Qutrit}

For the qutrit, the occupation probabilities are
\begin{equation}\label{eq.pj}
p_1 = \frac{e^{\beta \omega}}{Z} \,,\quad p_2 = \frac{1}{Z} \,,\quad p_3 = \frac{e^{-\beta \omega}}{Z} \,; \quad Z = 1+2\cosh\left(\beta \omega\right)\,,
\end{equation}
and from~(\ref{eq.pkpK}) one has $p_2^2 = p_1 \, p_3$. In the vicinity of the most mixed state ($\beta$ close to naught) behave as linear functions of the energy, i.e.,
\begin{equation}
p_1 \approx \frac{1}{3} ( 1 - \beta \, \, \omega) \, , \quad p_2 \approx \frac{1}{3} \, , \quad p_3 \approx \frac{1}{3} ( 1 + \beta \, \, \omega) \, .
\end{equation}

\begin{figure}
\centering
\includegraphics[width=0.4\linewidth]{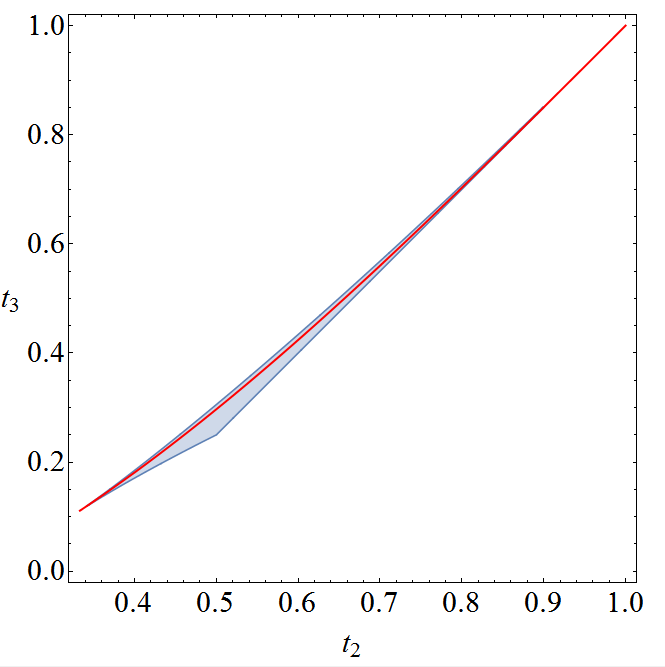}
\includegraphics[width=0.42\linewidth]{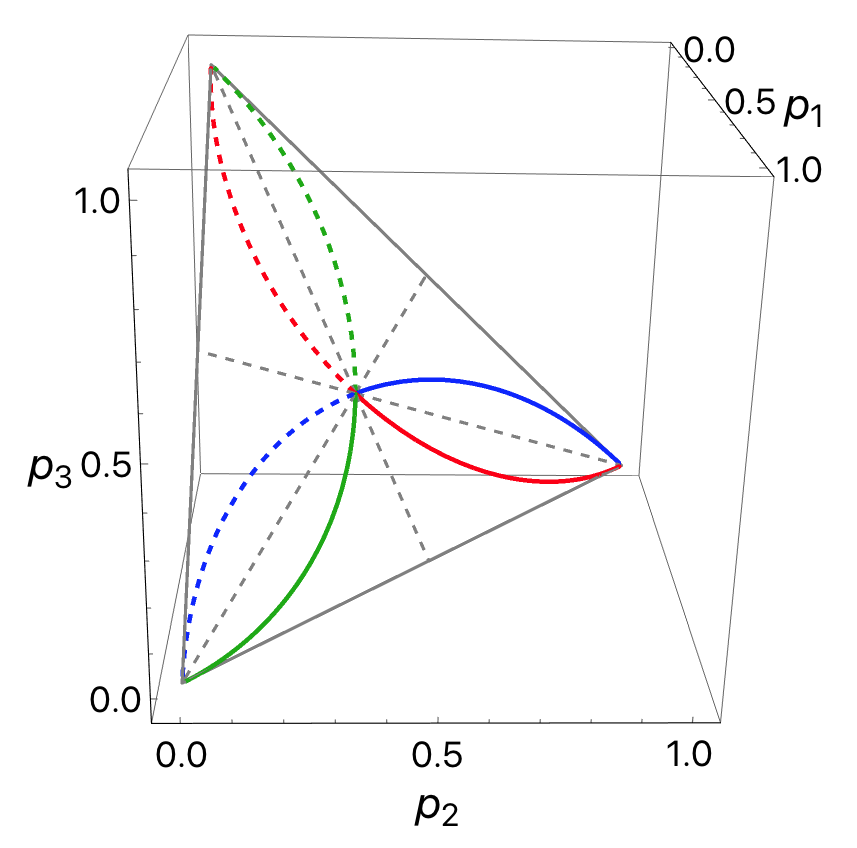}\\
\includegraphics[width=0.42\linewidth]{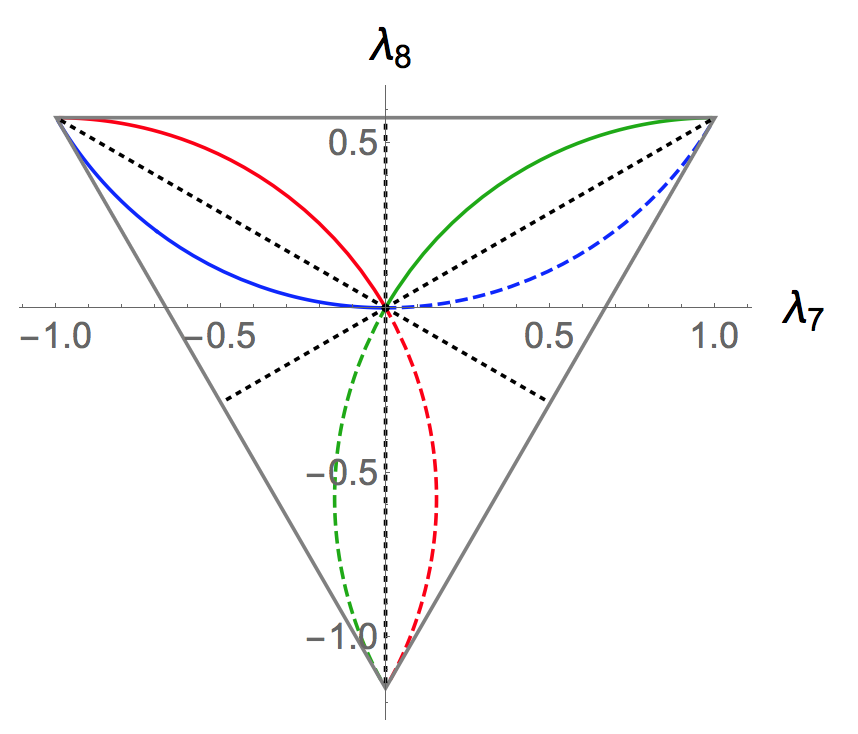} \quad
\includegraphics[scale=0.5]{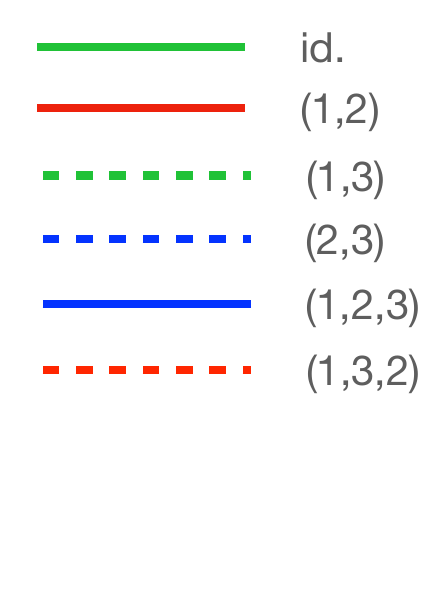}
\caption{Left: curve defined by the thermal states of a qutrit crossing the allowed physical region which defines a valid quantum system, from the most mixed state ($t_2=1/3$, $t_3=1/9$) ($T\rightarrow \infty$) to the pure state state ($t_2=t_3=1$) ($T\rightarrow 0$). Right: the same curve defined by the thermal states as seen in $p$-space follows a three-petal-flower-like structure. Bottom: The same flower-like structure is seen in $\lambda$-space. The legend shows the permutations of $(p_1,\,p_2,\,p_3)$ which lead to each flower-like petal.}
\label{3-thermal}
\end{figure}

The  invariants of the thermal density matrix are given as function of the temperature as
\begin{equation}
t_\ell  = \frac{1+2\cosh(\ell \beta \omega)}{Z^\ell}\,, \quad \ell = 1,2,3\,;
\end{equation}
a straightforward algebraic calculation shows that these are not independent, they satisfy
\begin{equation}\label{eq.t3t2thermal}
t_3 = \frac{1}{8} \, ( 9 \, t^2_2 - 3 \, t^3_2 +3 \, t_2 - 1) \, .
\end{equation}
This defines a curve [red line in Figure~\ref{3-thermal} (left)] which connects, as a function of the temperature, the most mixed state ($\beta\to 0$) to the pure state ($\beta \to \infty$) in the simplex, crossing the allowed quantum system region.

The map between the probability or Gell-Mann representations and the invariant representation is non-linear (cf. Section~\ref{ss.matrixR}), since each vertex of the simplex in $p$-space or $\lambda$-space represents a pure state, while all pure states fall on a single vertex in $t$-space.

The curve in $p$-space to which the thermal curve~(\ref{eq.t3t2thermal}) in $t$-space maps, depends on the particular election of the axes. For instance, if the occupation probabilities for the ground, first, and second states are labeled by $p_1,\, p_2$ and $p_3$ respectively, and a point in the simplex by $\vec{p}=(p_1,\,p_2,\,p_3)$, the thermal curve will connect the most mixed state $\vec{p}_e$ with the ground state $p_1=(1,0,0)$; for the choice of axes $\vec{q}=(p_1,\,p_3,\,p_2)$ the thermal curve suffers a specular reflection with respect to the bisectrix at vertex $\vec{p_1}$. Permutations $\vec{u} = (p_3,\,p_1,\,p_2)$ or $\vec{v} = (p_2,\,p_3,\,p_1)$ rotate the thermal curve by $2\pi/3$ and $4\pi/3$ counterclockwise respectively. All these permutations correspond physically to different orderings of the energy levels as functions of the Hamiltonian control parameters, they then characterize first order phase transitions in the system. Second order transitions do not involve a permutation of the simplex axes, they take place when the thermal state, as a function of the control parameters, either comes close to a bisectrix (non-touching avoided crossing) or touches the bisectrix (touching avoided crossing) (cf.~\cite{cordero21, lopez-pena21}).

In Figs.~\ref{f.simplexPL} and~\ref{f.simplex3L}(c) the black curve in $\lambda$-space remains when both $\omega$ or the temperature are changed. The state therefore remains confined within one sector of the corresponding simplex (i.e., it is subject to no phase transitions).
 
By considering all permutations, the thermal curve~(\ref{eq.t3t2thermal}) is maped to a three-petal flower-like structure in the probability representation, as shown in Figure~\ref{3-thermal} (right).  Since the map between $p$-space and $\lambda$-space is linear (Sec.~\ref{ss.matrixR}), the flower-like structure in $p$-space remains in $\lambda$-space, as shown in Figure~\ref{3-thermal} (bottom), where the legend shows the permutations leading to the structure.
%

\subsection*{Ququart}

\begin{figure}
\begin{center}
\includegraphics[scale=0.17]{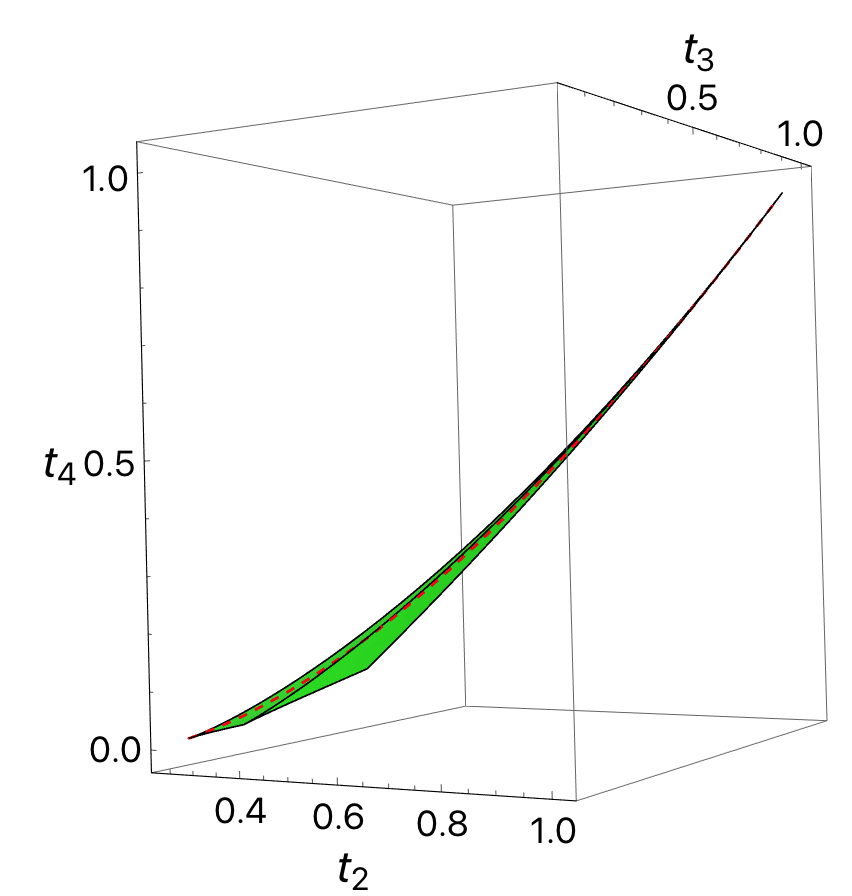}
\includegraphics[scale=0.185]{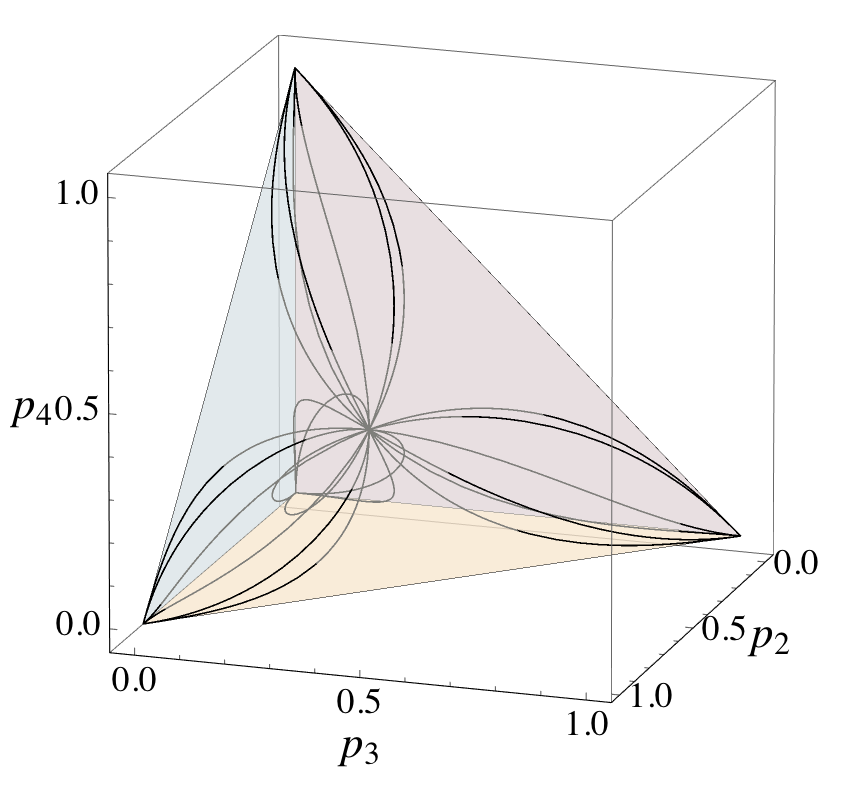}
\includegraphics[scale=0.3]{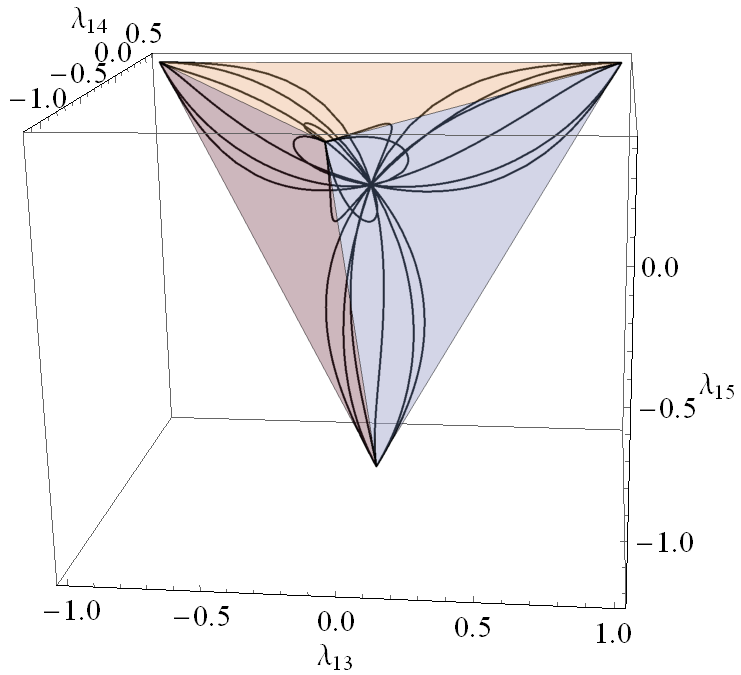}
\end{center}
\caption{Left: dashed dark (red) line defined by the thermal equilibrium states of a ququart, crossing the allowed physical region which defines a valid quantum system. This curve moves from the most mixed state (lower left, with $t_2=1/4$, $t_3=1/16$, $t_4=1/64$ and $T\rightarrow \infty$) to the pure state (upper right, with $t_2=t_3=t_4=1$ and $T\rightarrow 0$). Center: the same curve defined by the thermal states as seen in $p$-space follows a three-petal-flower-like structure; the most mixed state lies at the center of the simplex, with the pure states lying at the vertices; all permutations of the labeling of the states have been taken [for the relation between the $t$- and $p$-spaces see eqs.~(\ref{eq.vpr}) and  (\ref{p_to_t_b})]. Right: The same flower-like structure is seen in $\lambda$-space. See text for details.}
\label{fig9}
\end{figure}

For the ququart with $J=3/2$ in Eq.~(\ref{eq1b}), the occupation probabilities read as
\begin{eqnarray}
p_1&=& \frac{e^{3\beta \omega/2}}{Z}\,,\quad p_2= \frac{e^{\beta \omega/2}}{Z}\,,\quad p_3= \frac{e^{-\beta \omega/2}}{Z}\,,\quad p_4= \frac{e^{-3\beta \omega/2}}{Z}\,,\\[2mm] 
Z&=& 2\left[\cosh(\beta\omega/2)+\cosh(3\beta\omega/2)\right]\,, \nonumber
\end{eqnarray}
and the invariants take the form
\begin{equation}
t_\ell = 2\frac{\cosh(\ell \beta\omega/2)+\cosh(3\ell \beta\omega/2)}{Z^\ell}\,,\qquad \ell =1,\,2,\,3,\,4\,;
\end{equation}
this yields, in a parametric form $[t_2(\beta),\, t_3(\beta), t_4(\beta)]$, the thermal curve in the simplex, connecting, as a function of the temperature, the most mixed state $(t_2,t_3,t_4)=(1/4,\, 1/16,\, 1/64)$ with the vertex $(t_2,t_3,t_4)=(1,\,1,\,1)$ corresponding to the pure ground state. This is shown in figure~\ref{fig9}(left) [red line]. In a similar manner to the qutrit case above, by considering the different permutations for the election of the axes in the probability or the Gell-Mann representation (in this case $24$), the thermal curve in the invariant representation is mapped to a flower-like structure  as shown in figure~\ref{fig9} (center) [black lines] for the probability representation, which remains in the Gell-Mann representation Fig.~\ref{fig9} (right) [black lines].

For the diagonal Hamiltonian~(\ref{eq.Hdiag}) the occupation probabilities of the thermal state satisfy $p_j > p_k$ for $j < k$ Eq.~(\ref{eq1a}), since the spectrum energy values Eq.~(\ref{eq.EM}) satisfy the same relationship. Hence, for a fixed axes ordering of the probabilities, the thermal curve remains inside one sector delimited by bisectrices [cf. Fig.~\ref{f.simplexPL} and Fig.~\ref{f.simplex3L}(c)], i.e., there are no quantum phase transitions. Phase transitions mirrored in the simplex require non linear Hamiltonians.

\subsection{Lipkin-Meshkov-Glick model}

The LMG model is a many body quantum system used in several fields like nuclear physics~\cite{LMG65a, LMG65b, LMG65c}, condensed matter~\cite{botet82, botet83, citlali10} and quantum optics~\cite{kitagawa93}, which is written in terms of collective operators in the angular momentum algebra. 

\subsection*{Qutrit} 

\begin{figure}
\begin{center}
\includegraphics[width=0.4\linewidth]{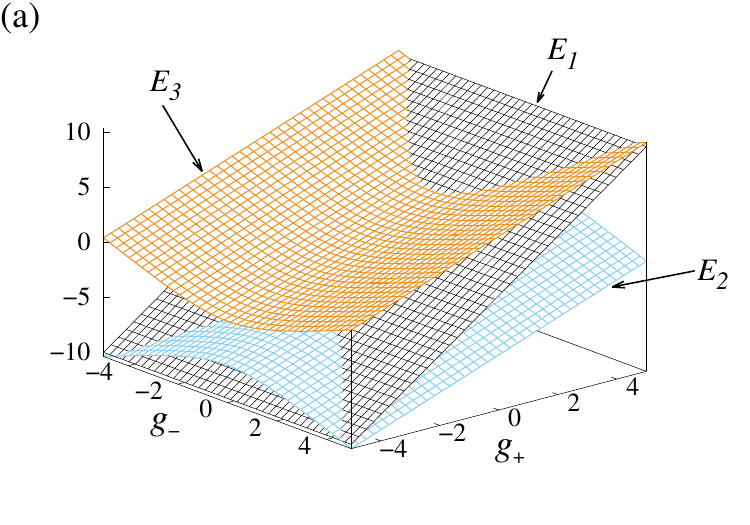} \quad
\includegraphics[width=0.4\linewidth]{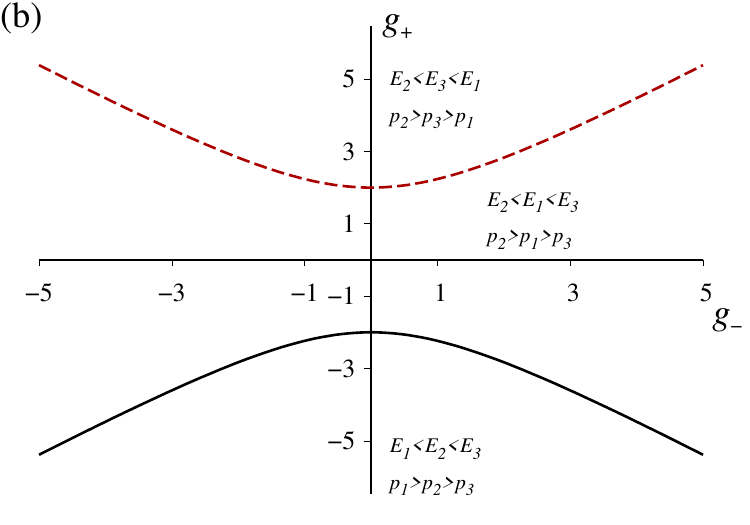} \\
\includegraphics[width=0.4\linewidth]{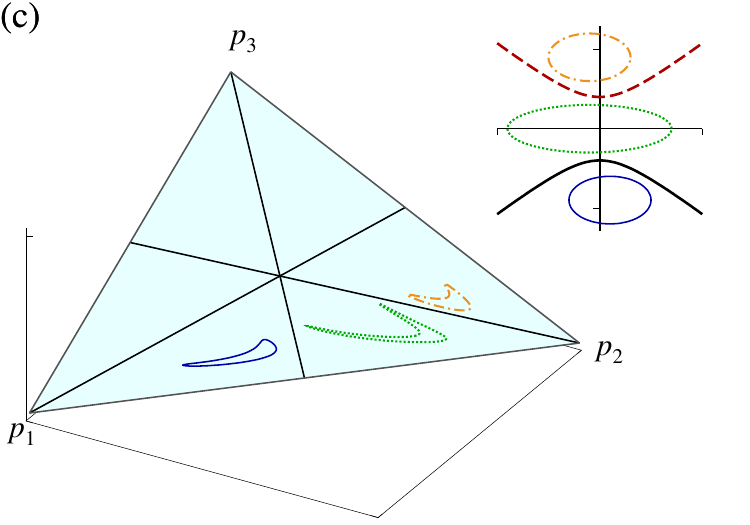} \quad 
\includegraphics[width=0.4\linewidth]{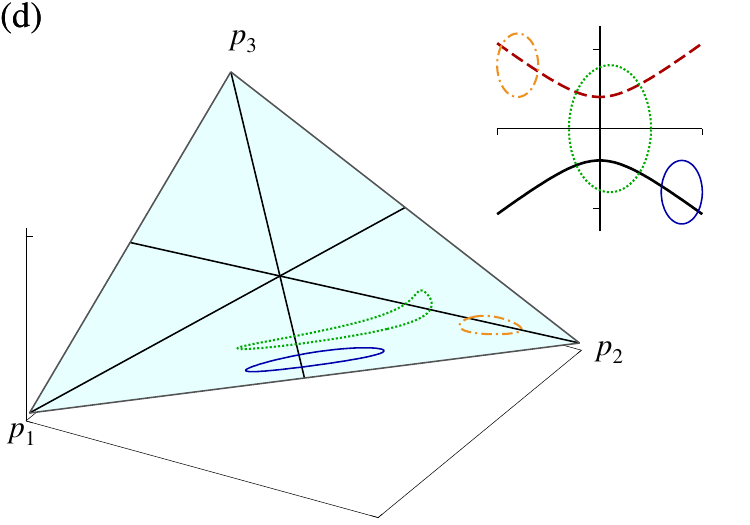}\\
\includegraphics[width=0.4\linewidth]{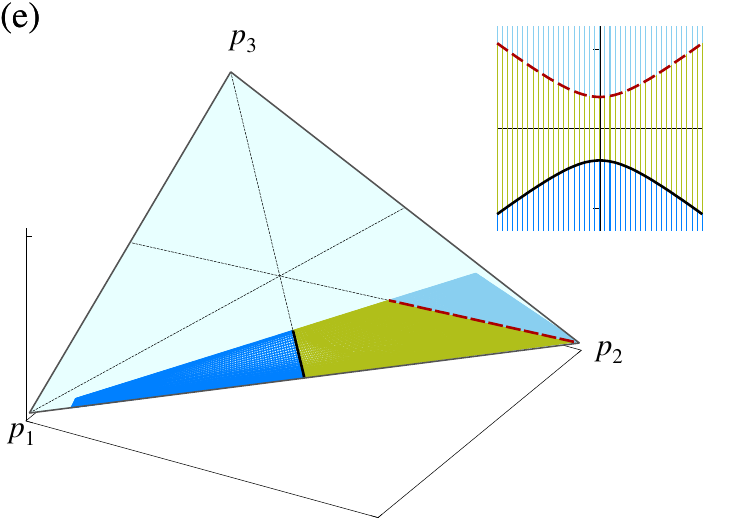}
\end{center}
\caption{LGM model with $J=1$ and dimensionless parameters: (a) Energy surfaces in units of $\omega$; (b) energy separatrices for $E_1=E_2$ (solid black line)  and for  $E_1=E_3$  (dashed red line), indicating the corresponding energy and probability conditions for thermal states. (c) Continuous, dotted, and dash-dot lines corresponding to thermal states with $\beta \, \omega=1/4$ without transitions in the $p$-space simplex, and (d) same as (c) for states that suffer at least one quantum phase transition (shown in the corresponding insets). (e) Map onto the simplex of the different regions in parameter space.}
\label{f.energyl3}
\end{figure}

The LGM Hamiltonian for $N=2$ particles ($J=1$),  given in units of $2\,\omega$ is 
\begin{equation}
\frac{1}{2\omega}\,\op{H} =  \op{J}_z + g_x \, \op{J}^2_x + g_y \, \op{J}^2_y \, ,
\end{equation}
whose energy eigenvalues are denoted by
\begin{equation}
E_1 =  2 \, \omega \, g_+\,,\qquad
E_2 =  \omega\left[ \, g_+ - \sqrt{4+g_-^2}\right]\,,\qquad
E_3 =  \omega\left[ g_+ + \sqrt{4+g_-^2}\right]\, ,
\label{eq.energyl3}
\end{equation}
which determine the corresponding occupation probabilities $p_i=e^{-\beta E_i}/Z$ of the thermal state as functions of the  dimensionless control parameters $g_\pm=g_x\pm g_y$. The plots of these energy surfaces are shown in figure~\ref{f.energyl3}(a); crossings of the energy levels show first order transitions in both the ground and the most excited states. By imposing the conditions $E_1=E_2$ and $E_1=E_3$ one obtains, respectively, the energy separatrices
\begin{equation*}
g_+ = - \sqrt{4+g_-^2}\,;\qquad  g_+ =  \sqrt{4+g_-^2}\, . 
\end{equation*}
%

These divide  the control parameter space in three regions: (I) points below the separatrix for the ground state $g_+ = - \sqrt{4+g_-^2}$ correspond to the order $E_1<E_2<E_3$ giving rise to the probability ordering $p_1>p_2>p_3$, (II) points between the separatrices $g_+ = \pm \sqrt{4+g_-^2}$ correspond to a spectrum with  $E_2<E_1<E_3$ implying that $p_2>p_1>p_3$, and (III) the points above the separatrix for excited states $g_+ = \sqrt{4+g_-^2}$ where $E_2<E_3<E_1$ with $p_2>p_3>p_1$, as shown in figure~\ref{f.energyl3}(b). 

From the relation between the energies and probabilities for thermal states, in a similar manner the simplex in probability or Gell-Mann representations divides itself into different regions.  These different energy regions may be seen as permutations of the occupation probabilities leading to the following transitions: from region I to II or viceversa ($E_1\rightleftharpoons E_2$) one has a ground state transition, which is obtained in $p$-space by means of a mirror transition about the bisectrix of the vertex $\vec{p}_3=(0,0,1)$; from region II to III or viceversa ($E_1\rightleftharpoons E_3$) a transition between excited states takes place, with $E_2$ for the ground state, obtained by a mirror transformation in the simplex with respect to the bisectrix at the vertex $\vec{p}_2=(0,1,0)$. 

By fixing the energy labels Eq.~(\ref{eq.energyl3}), the vertices of the simplex are fixed.  Thus, the thermal states map the control parameter space to the simplex: for $\beta\to 0$ the parameter space coalesces to the most mixed state $\vec{p}_e=(1/3,1/3,1/3)$, while for $\beta\to\infty$ region I is mapped to the vertex $\vec{p}_1=(1,0,0)$, regions II and III to the vertex $\vec{p}_2=(0,1,0)$, the separatrix of the ground state maps to the point $\vec{p}_{m1}=(1/2,1/2,0)$ and that of the most excited state coalesces at  $\vec{p}_{m2}=(0,1/2,1/2)$. For other cases, finite non-zero temperature, different regions of parameter space map to different regions in the simplex.

In figure~\ref{f.energyl3}(c), paths inside the regions of the energy phase diagram I (solid line), II (dotted line), and III (dash-dot line), are mapped to the simplex as function of the control parameters ($g_+, g_-$). No quantum phase transitions are present, i.e., all points remain in one sector of the simplex. On the other hand, figure~\ref{f.energyl3}(d) shows the map when at least one transition occurs; the points on the quantum phase diagram are shown in the inset. The solid (blue) trajectory moves from a sector where $\vec{p}_1$ represents the ground state to a region where $\vec{p}_2$ represents the ground state; this may be seen as a reflection of the simplex about the bisectrix at vertex $\vec{p}_3$. A similar behavior occurs for the excited state transition (dash-dot, orange line) where the reflection now takes place with respect to the bisectrix at vertex $\vec{p}_2$. The dotted (green) line, which crosses both transitions, shows all the posible reflections. As expected, the points on the separatrix are located along a bisectrix, with either $E_1=E_2$  or $E_1=E_3$, as can be seen in figure~\ref{f.energyl3}(e).

\begin{figure}
\centering
\includegraphics[width=0.4\linewidth]{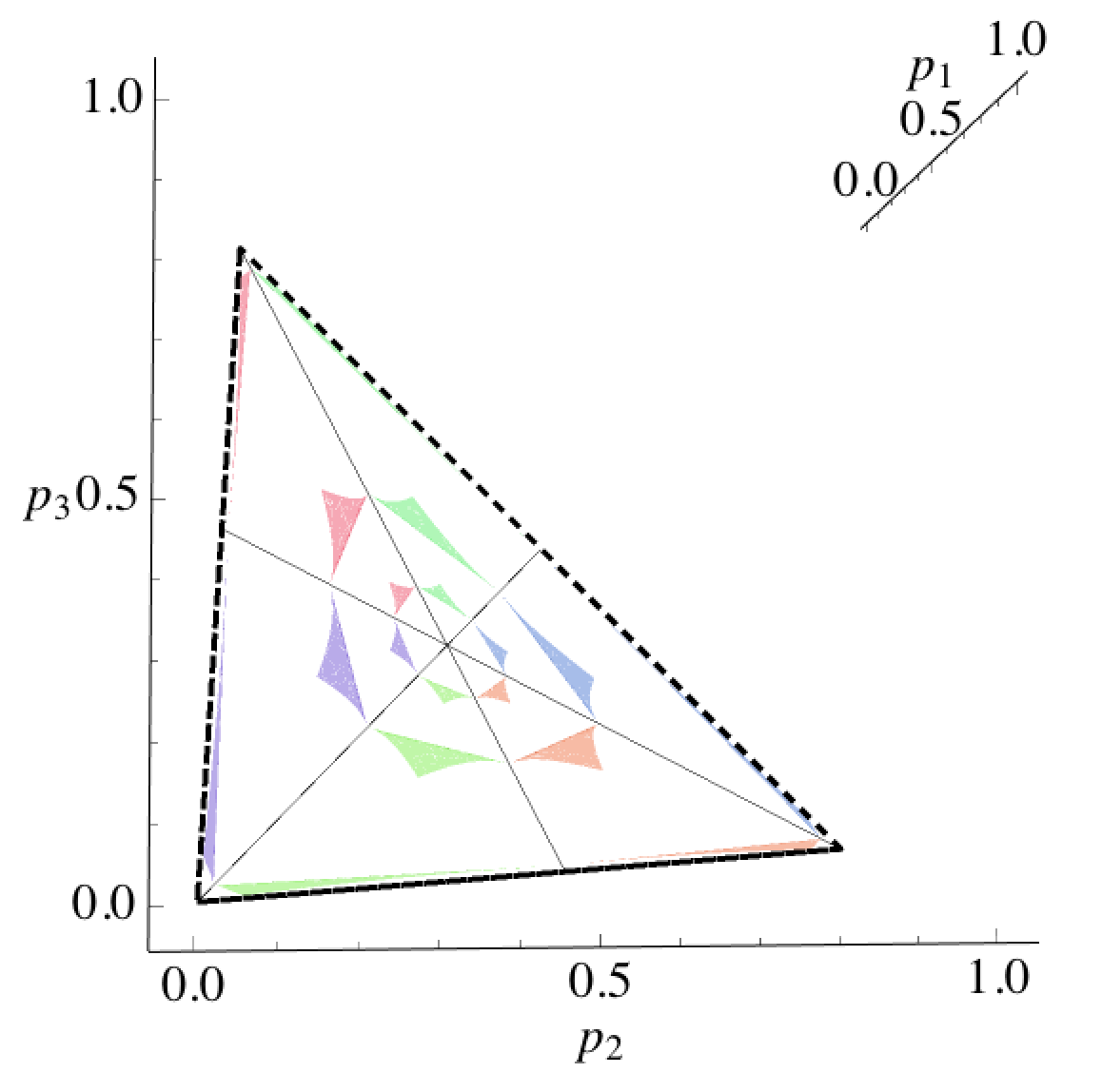} \quad
\includegraphics[width=0.4\linewidth]{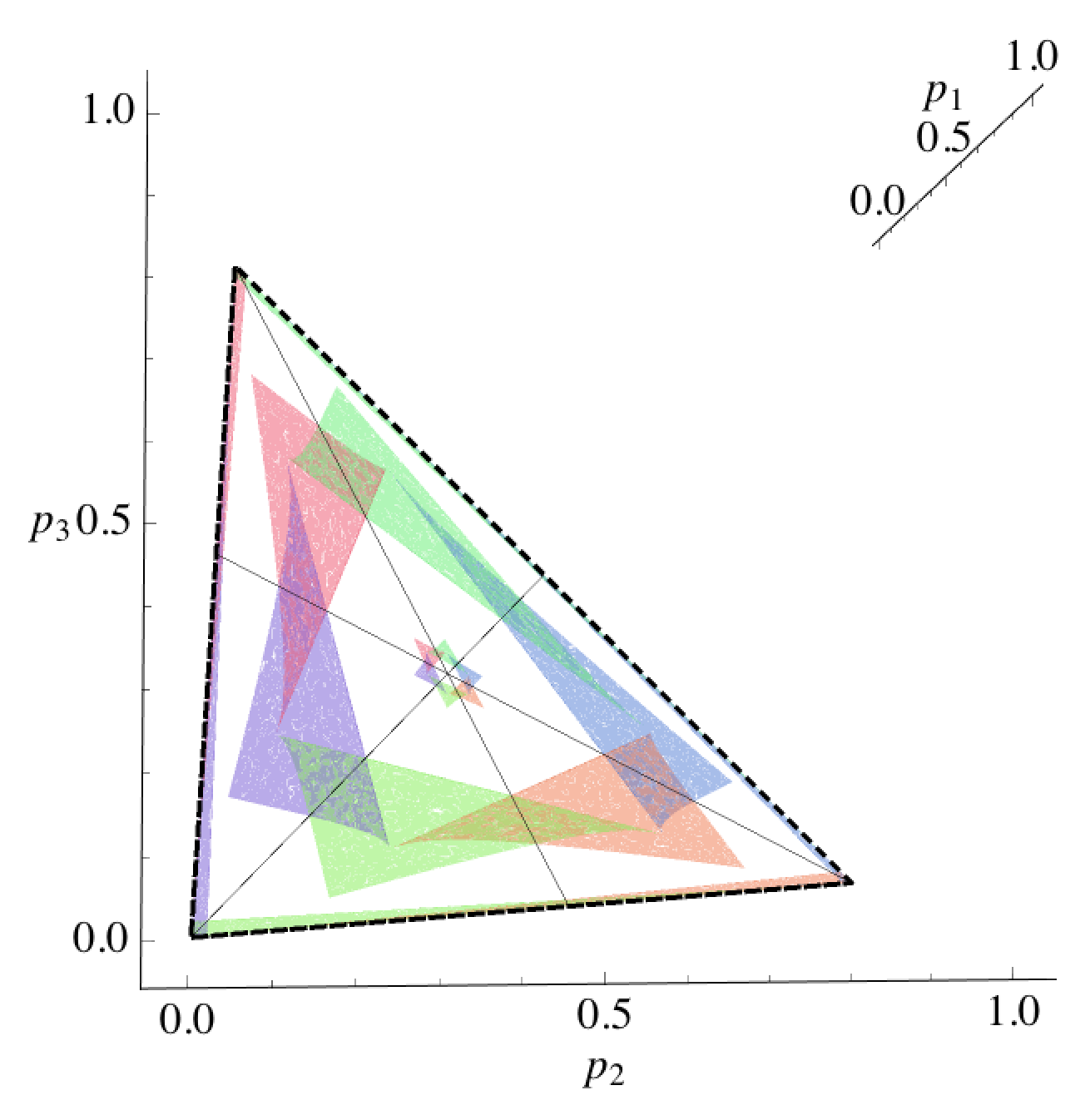}
\caption{Left: Probabilities for the three-level diagonal thermal equilibrium state defined by the LMG Hamiltonian model, for $\beta=1/5$ (center image), $\beta=1/2$ (middle image), and $\beta = 2$ (outer image). As $\beta$ increases (temperature decreases) the allowed zones approach the edges of the simplex, corresponding to pure states. Note that all the allowed regions remain within one of the sectors in the simplex, no phase transitions take place. In all the cases $j=1$ (qutrit) and $g_{\pm} \in (-2,2)$. Right: Same as before, for $g_{\pm} \in (-6,6)$ and with $\beta=1/20$ (center image), $\beta=1/2$ (middle image), and $\beta = 2$ (outer image). By increasing the values of the coupling parameters, allowed regions overlap due to the inversion of the atomic level ordering.}
\label{fig11}
\end{figure}

We have therefore two cases: the case when there are no quantum phase transitions displayed in Fig.~\ref{fig11} (left) and the case when there are quantum phase transitions exhibited in Fig.~\ref{fig11} (right). In these figures one considers all the possible permutations of the symmetric group of three elements. The first of these shows in $p$-space the behavior of the occupancy probabilities when the parameter $\beta$ takes the set of values $\{ 1/5, 1/2, 2\}$. The regions expand outwards towards the boundaries of the triangle as $\beta$ increases when the parameters of the Hamiltonian model are within the interval $g_{\pm} \in (-2,2)$. The second case (right) shows in $p$-space the occupancy probabilities for the parameter $\beta$ taking the values $\{ 1/120, 1/2, 2\}$. Again the regions move from the geometrical center to the boundaries of the triangle as $\beta$ increases. The indicated transitions are associated to the interchange of the occupancy probabilities $(p_1,p_2), (p_1,p_3), (p_2,p_3)$. In this case the control parameters of the Hamiltonian vary within the interval $g_{\pm} \in (-6,6)$.

\subsection*{Ququart}

The LMG Hamiltonian model for $N=3$ particles ($J=3/2$) represents a ququart in quantum information theory, with energy levels given by
\begin{eqnarray*}
E_1 &=& \frac{\omega}{2}\left(5 g_+ + 2 - 2\sqrt{3g_-^2+ (g_+-2)^2}\right)\,,\quad 
E_2 = \frac{\omega}{2}\left(5 g_+ - 2 - 2\sqrt{3g_-^2+ (g_++2)^2}\right)\,, \\[2mm]
E_3 &=& \frac{\omega}{2}\left(5 g_+ - 2 + 2\sqrt{3g_-^2+ (g_++2)^2}\right)\,,\quad
E_4 = \frac{\omega}{2}\left(5 g_+ + 2 + 2\sqrt{3g_-^2+ (g_+-2)^2}\right)\, ,
\end{eqnarray*}
whose occupation probabilities are indicated by $p_j=e^{-\beta E_j}/Z$ for the thermal state. 

\begin{figure}
\begin{center}
\includegraphics[width=0.4\linewidth]{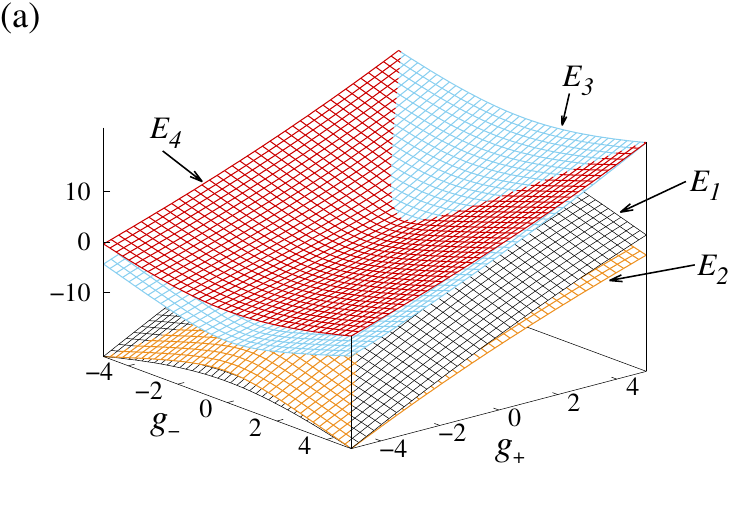}\quad 
\includegraphics[width=0.4\linewidth]{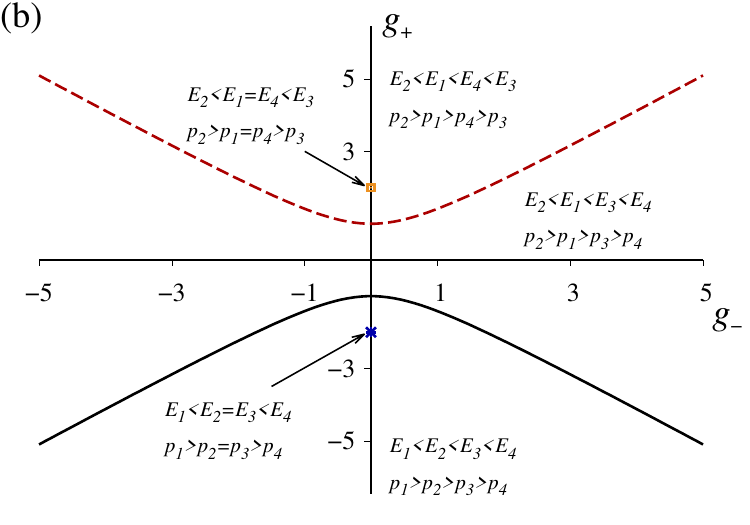}\\
\includegraphics[width=0.4\linewidth]{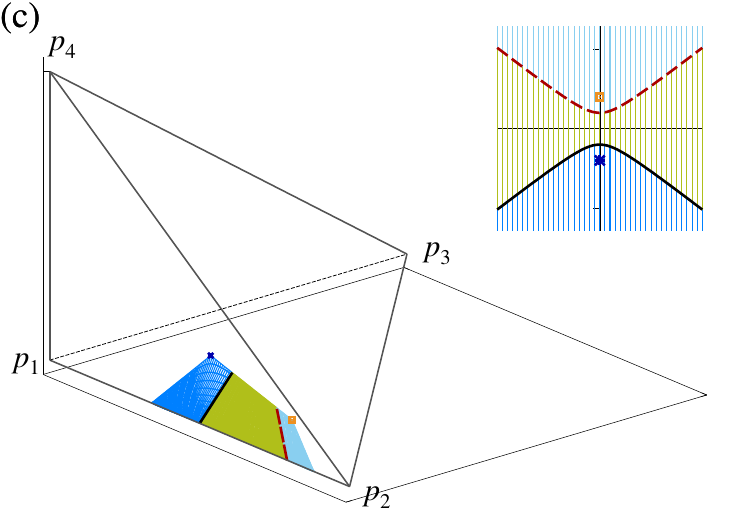}\quad
\includegraphics[width=0.4\linewidth]{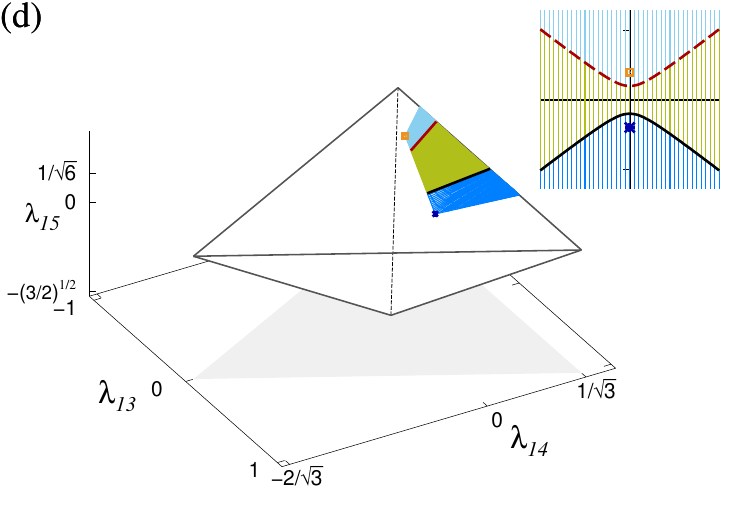}
\end{center}
\caption{
LGM model with $J=3/2$, in dimensionless parameters: (a) Energy surfaces in units of $\omega$; (b) separatrices for $E_1=E_2$ (solid black line) and for  $E_1=E_4$  (dashed red line); the point where $E_2=E_3$ is marked by an asterisk, while the point $E_1=E_4$ is marked by a square. The energy conditions for each of the three regions are indicated, together with the corresponding probability conditions for a thermal state. The map for $\beta\omega=1/3$ of the control parameter space onto the simplex is shown in (c) for the probability space $(p_2,p_3,p_4)$ (with $p_1=(0,0,0)$), and in (d) for the Gell-Mann space.}
\label{f.energyl4}
\end{figure}

The energy surfaces as functions of the control parameters $g_\pm$ are shown in Figure~\ref{f.energyl4}(a). Two first order transitions, one of these in the ground state where $E_1\rightleftharpoons E_2$ and the other associated to the excited states $E_3\rightleftharpoons E_4$, are shown. The separatrices are determined by the conditions $E_1=E_2$, for which $g_+ = - \sqrt{1+g_-^2}$,  and $E_3=E_4$, for which $g_+ =  \sqrt{1+g_-^2}$, and they divide the control parameter space in three regions.

At $(g_-,g_+)=(0, \pm2)$ there is a double degeneracy of the second and third excited states $(E_1=E_4)$ and $(E_2=E_3)$, respectively. The  quantum phase diagram is shown in figure~\ref{f.energyl4}(b), showing region (I) for $E_1<E_2<E_3<E_4$ with $p_1>p_2>p_3>p_4$, region (II) for $E_2<E_1<E_3<E_4$ with $p_2>p_1>p_3>p_4$ and region (III) for $E_2<E_1<E_4<E_3$ with $p_2>p_1>p_4>p_3$. The solid (black) line shows the separatrix of the ground state, while the dashed (red) line shows that of the excited states;  the point where $E_2=E_3$ in region I is indicated with an asterisk, and for $E_1=E_4$ with an square. The simplex in this case is a $3$-dimensional tetrahedron in a space of four dimensions, and may be represented via three-dimensional slices as in figures~\ref{f.simplex3P} (or \ref{f.simplex3L}) for the probability (or Gell-Mann) representation. Each vertex of the tetrahedron represents a pure state and, for a fixed selection of these, the points of a thermal state remain only in one sector [cf. Fig.~\ref{f.simplex3L}(c)]; thus, each region in parameter space is mapped onto only one sector of the tetrahedron. The map of the control parameter space onto the simplex via the thermal state for the value $\beta\omega=1/3$ is shown in Fig.~\ref{f.energyl4}(c) for the probability representation, and in Fig.~\ref{f.energyl4}(d) for Gell-Mann representation.

In the four-level system we have a similar behavior as in the qutrit case. Here the zones where the probabilities follow a certain order (say $p_1 > p_2 > p_3 > p_4$) are delimited by one of 24 different tetrahedrons inside the 3-simplex of Figure~\ref{f.simplex2P}. Any crossing between these 24 tetrahedrons means that the there is a quantum phase transition (for example $p_1 > p_2 > p_3 > p_4$ changes to $p_2 > p_1 > p_3 > p_4$ in two different tetrahedrons).

In the case of the LMG model, different behavior (non-crossing versus crossing of different zones) is observed when the values of two probabilities are equal, i.e., when the control parameters lie on the separatrix $g^2_+ -g^2_{-}=1$. Figure~\ref{f.simplex3L}(b) shows a  plot for the cases:  $p_1=p_2$ (degeneracy of the ground state)  and $p_3=p_4$ (degeneracy of the most excited states.

Similarly to the three-level system, as $T\to\infty$ we get the most mixed state at the center of the tetrahedron, while for $T \to 0$ the possible states approach the vertices of the simplex. Taking values of the coupling parameters based on the energy phase diagrams shown in Fig.~\ref{f.energyl4} one may obtain crossings between the different regions, physically corresponding to an inversion of population.

\section{Summary and concluding remarks}
\label{s-conclusions}

The PME yields the Boltzmann probability distribution function for any Hamiltonian system characterizing an ensemble of particles with fixed average energy, and one is able to maximize the von Neumann entropy and minimize the Helmholtz free energy. In this work we have considered the probability ($2$- and $3$-simplices), the Gell-Mann, and the invariant spaces. In all of them, one is able to visualize the purity properties of  qutrit and ququart systems: i) where there is maximum mixture, ii) the regions with eigenvalue degeneracies, and iii) the loci where all the eigenvalues are different. The linear and non-linear maps between the probability, Gell-Mann, and invariants representations were given, which additionally allow us to determine general geometric properties of the qutrit and ququart systems such as: i) trajectories in the $p$-, $\lambda$-, and $t$- spaces obtained by changing the temperature associated to the density matrix, and in this form visualize immediately the corresponding properties of the considered state; ii) trajectories associated to fixed values of the invariants $t_2$, $t_3$, and $t_4$ (see Figs.~\ref{fig1}, \ref{fig2}, \ref{f.simplex3P}, \ref{f.simplex3L}).  

A relation between the order of the quantum transitions for the discrete spectrum, and rotations of the simplex for the thermal state in both, probabilistic and Gell-Mann representations, was given via rotations and reflections (for first order transitions) or when the path in the simplex approaches a bisectrix (for second order transitions). This was illustrated for two Hamiltonian models: (a) linear Hamiltonians, which present no transitions, and in this case the thermal curve on the simplex remains inside a sector of the simplex, and (b) non-linear Hamiltonians, where transitions of the ground and excited states occur; the effect on the simplex of the quantum transitions in the probability representation was shown for the qutrit Fig.~\ref{f.energyl3} and the ququart Fig.~\ref{f.energyl4}. In this case, the connection between the probability space and the energy phase diagram is direct. In other systems this connection is more complex, but it is still illuminating to study the simplex geometry as was done in~\cite{cordero21, lopez-pena21} for three-level atomic systems in the presence of a radiation field, for different configurations. It will also be interesting to carry this study to other systems.

Thermal states associated to the linear Hamiltonian in the angular momentum operators were studied. For the qutrit, the curve $p^2_j= p_i p_k$ (with $p_i\leq p_j\leq p_k$ in order to satisfy the energy conditions) and their corresponding permutations give rise to three-petal-flower-like structure in the $2$-simplex, both in $p$-space and $\lambda$-space. When the Hamiltonian interaction strengths $g_\pm$ describe hyperbolae with axes equal to $2 \omega$, we have double eigenvalue degeneracy. For the ququart, from Eq.~(\ref{eq.pkpK}), the curve $p_i p_\ell = p_j p_k$ (with $p_i\leq p_j\leq p_k\leq p_\ell$ in order to satisfy the energy conditions) and its corresponding permutations, similar geometric structures were given in its $3$-simplex (the tetrahedron).

The analysis of the quadratic LMG Hamiltonian model includes the qutrit $J=1$ and the ququart $J=3/2$. Very different trajectories are obtained for the same temperatures $\beta=1/20,\, 1/2,\, 2$ when constraining the values of $g_{\pm}$ in the interval $[-2,2]$, in contrast to the paths when $g_{\pm}$ lie in the interval $[-6,6]$ (see Fig.~\ref{fig11}). For $\beta=1/20,\, 1/2,\, 2$ the trajectories in $p$-space are shown for $g_{\pm}$ in the interval $[-2,2]$, and in both cases (qutrit and ququart) they move towards the boundary of the simplex as $\beta$ increases (temperature decreases).

The results above provide a much clearer understanding of the geometry properties of qudit systems in general, and present specific results for qutrit and ququart systems.

It would be interesting to determine the behavior of the spin squeezing of the LMG model in the different simplex representations studied here. Spin squeezing has been extended to describe general multi-qudit systems~\cite{calixto21, guerrero22}. Embedding of the $\mathfrak{su}(2)$ sub-algebras into $\mathfrak{u}(n)$ yields $n(n-1)/2$ squeezing parameters. It is known that these, together with entanglement, are good markers of the quantum phase transitions and localization properties in phase space exhibited by the LMG Hamiltonian system of $n$-level interacting atoms~\cite{mayorgas23}.

\appendix
\section{Gell-Mann Parametrization}
\label{ap1}
The density matrix description of a state of $n$-level systems requires $n^2-1$ real parameters is given by~(\ref{eq.rhoL}), that is
\begin{equation}
\op \rho = \frac{1}{n} {\op I} +  \frac{1}{2} \sum_{k=1}^{n^2-1} \lambda_k \, \op \Lambda_k \, ,
\label{matriz}
\end{equation}
where ${\op I}$ is the identity matrix in $n$ dimensions and $\lambda_k$ is the expectation value of the generator $\op{\Lambda}_k$. The operator $\op \rho$ is hermitian and satisfies ${\rm Tr}(\op \rho) =1$ because the generators $\op\Lambda_k$ of the special unitary algebra $\mathfrak{su}(n)$ are complex traceless hermitian matrices. These generators have the property ${\rm Tr}({\op{\Lambda}_j} \op{\Lambda}_k)=~2 \, \delta_{j \, k}$.  

The important properties for the description of the density matrix state are the positivity condition $\op{\rho}_k \geq 0$ and the fact that $1/n \leq {\rm Tr}\,(\op{\rho}^2)\leq 1$, which are fundamental to determine the physical space of the density matrices~\cite{procesi78, procesi07}. Notice that the generators are determined by means of the symmetric $d_{ijk}$ and antisymmetric $f_{ijk}$ structure constants of the mentioned special unitary algebra. The condition, ${\rm Tr}\,(\op{\rho}^2)\leq~1$ on~(\ref{matriz}) implies that the length of the Bloch vector is bounded, that is, 
\begin{equation}
\vert \lambda\vert \leq \sqrt{\frac{2 \, (n-1)}{n}} \, .
\end{equation}
The density matrix~(\ref{matriz}) satisfies the positivity condition if its characteristic polynomial $p_\rho(x)=\det(x {\op I} -\op{\rho})$, viz.
\begin{equation}
p_\rho(x)= x^n - a_{1} \, x^{n-1} + a_{2}\,  x^{n-2}- \cdots + (-1)^{n-1} \, a_{n-1} \, x + (-1)^n \, a_n \, ,
\end{equation}
has coefficients $a_k \geq 0$ according to the Descartes theorem about the sign rules~\cite{curtiss18}.

An additional remark is that the orbits generated by unitary transformations on the density matrices can be classified in terms of the Gram matrix, built from the Hilbert-Schmidt scalar products of the vectors lying in the tangent space of the orbits, and the rank of the Gram matrix determines the dimensions of the orbits. This classification is consistent with the results obtained by means of the stability groups of unitary transformations~\cite{arvind97, khanna97}. Due to the invariance of the Gram matrix under unitary transformations, in order to determine the orbits one can consider the diagonal representation of the density matrix~\cite{kus01, bengtsson06, tay08, gerdt14}.

\section{Generators and simplex parametrization of unitary algebras in n dimensions}

A special unitary algebra in $n$ dimensions can be denoted by a set of operators $\op{\Lambda}_k$ with $k=1,2,\cdots n^2-1$. They are traceless ${\rm Tr}( \op{\Lambda}_k) =0$ and satisfy ${\rm Tr}(\op{\Lambda}_j \op \Lambda_k)= 2 \delta_{jk}$. Furthermore, these operators can be generated by the symmetric and antisymmetric linear combinations~\cite{kimura03,bruning12},
\begin{equation}
\op{A}_{jk}+\op A_{kj} \, ,\quad  - i (\op A_{jk}- \op A_{kj} ) \, , \quad j<k.
\end{equation}
together with the diagonal matrices defined by
\begin{equation}
(\op{F}_\ell)_{rr} = \sqrt{\frac{2}{\ell(\ell+1)}}\left\{\begin{array}{c l} 1 & r\leq\ell, \\ -\ell & r=\ell+1, \\ 0 & r>\ell+1, \end{array}\right.
\label{eq.Fellrr}
\end{equation}
with $\ell=1,2,\cdots, n-1$. Typically they are ordered, placing first the symmetric $n(n-1)/2$, then the antisymmetric $n(n-1)/2$, and at the end the diagonal ($n-1$) matrices. In this order, the diagonal Gell-Mann matrices are labeled as
\begin{equation}\label{eq.LambdaD}
\op{\Lambda}_{k_\ell} := \op{F}_\ell\,;\qquad k_\ell:=n^2-n+\ell\,,\quad \ell=1,\,2,\,\dots,\,n-1\,.
\end{equation}

\subsection{Gell-Mann Parametrization}
\label{ap1}
The density matrix description of a state of $n$-level systems requires $n^2-1$ real parameters and it is given by~(\ref{eq.rhoL}), that is,
\begin{equation*}
\op \rho = \frac{1}{n} {\op I} +  \frac{1}{2} \sum_{k=1}^{n^2-1} \lambda_k \, \op \Lambda_k \,;\qquad {\rm Tr}\,[\op{\Lambda}_j]=0\,;\quad {\rm Tr}\,[\op{\Lambda}_j\op{\Lambda}_k]=2\delta_{jk} \,,
\end{equation*}
where ${\op I}$ is the identity matrix in $n$ dimensions and $\lambda_k$ is the expectation value of the generator $\op{\Lambda}_k$. The operator $\op \rho$ is hermitian and satisfies ${\rm Tr}(\op \rho) =1$ because the generators $\op\Lambda_k$ of the special unitary algebra $\mathfrak{su}(n)$ are complex traceless hermitian matrices. 

The important properties for the description of a density matrix state are the positivity condition $\op{\rho}_k \geq 0$ and the fact that $1/n \leq {\rm Tr}\,(\op{\rho}^2)\leq 1$, which are fundamental to determine its physical space~\cite{procesi78, procesi07}. The generators are determined by means of the symmetric $d_{ijk}$ and antisymmetric $f_{ijk}$ structure constants of the mentioned special unitary algebra. The condition, ${\rm Tr}\,(\op{\rho}^2)\leq~1$ on~(\ref{matriz}) implies that the length of the Bloch vector is bounded by
\begin{equation}
\vert \lambda\vert \leq \sqrt{\frac{2 \, (n-1)}{n}} \, .
\end{equation}
The density matrix~(\ref{matriz}) satisfies the positivity condition if its characteristic polynomial $p_\rho(x)=\det(x {\op I} -\op{\rho})$, viz.
\begin{equation}
p_\rho(x)= x^n - a_{1} \, x^{n-1} + a_{2}\,  x^{n-2}- \cdots + (-1)^{n-1} \, a_{n-1} \, x + (-1)^n \, a_n \, ,
\end{equation}
has coefficients $a_k \geq 0$ according to the Descartes theorem about the sign rules~\cite{curtiss18}.

As an additional remark, the orbits generated by unitary transformations on the density matrices can be classified in terms of the Gram matrix, built from the Hilbert-Schmidt scalar products of the vectors lying in the tangent space of the orbits, and the rank of the Gram matrix determines the dimensions of the orbits. This classification is consistent with the results obtained by means of the stability groups of unitary transformations~\cite{arvind97, khanna97}. Due to the invariance of the Gram matrix under unitary transformations, in order to determine the orbits one can consider the diagonal representation of the density matrix~\cite{kus01, bengtsson06, tay08, gerdt14}.

\subsection{Simplex probability parametrization}\label{ap.pvectors}

Equation~(\ref{matriz}) has in a simple way a geometric representation of a point in the simplex of the probability space ($p$-space) of  dimension $n$, in terms of the orthonormal vectors formed from the diagonal elements of $\op{F}_\ell$
\[
\hat{e}_\ell := \frac{1}{\sqrt{2}} {\rm Diag}\,(\op{F}_{\ell}) =  \frac{1}{\sqrt{2}} \left([\op{F}_{\ell}]_{11},\,[\op{F}_{\ell}]_{22},\,\dots\,,[\op{F}_{\ell}]_{nn} \right)\,,
\]
and the most mixed state from the identity matrix $\op{I}$
\[
\vec{p}_e = \frac{1}{n} {\rm Diag}\,(\op{I}) = \frac{1}{n}\left(1,\,1,\,\dots\,,1\right)\, ,
\]
which satisfies the constraint $\sum^n_{k=1} {p}_{e k} =1$.
Then the simplex is defined as the $n$ equidistant points of the vector with maximum entanglement. These orthonormal vectors are given by~[eq.~(\ref{eq.LambdaD})],
\begin{equation}\label{eq.eellk}
[\hat{e}_\ell]_{k} = \frac{1}{\sqrt{\ell(\ell+1)}}\left\{\begin{array}{c l} 1 & k\leq\ell \\ -\ell & k=\ell+1 \\ 0 & k>\ell+1 \end{array}\right.\,,\quad k=1,\,2\,\dots,\, n\,,
\end{equation}
and $\ell =1,\, 2\, \dots n-1$, thus an arbitrary point in the simplex representation $\vec{p}$ is 
\begin{equation}
\vec{p}= \vec{p}_e + \sum_{\ell =1}^{n-1} a_\ell \hat{e}_{\ell}\,,
\end{equation}
where the constrictions in the coefficients $a_\ell$ are given by the maximum and minimum values of the invariants, $\frac{1}{n^{j-1}} \leq {\rm Tr}[\op{\rho}^j]\leq 1$ for $j=1,2,\dots,n$. Thus, the corresponding density matrix is determined by $\op{\rho}={\rm DiagM}\,(\vec{p})$, a diagonal matrix formed by the different probabilities (e.g., occupation probabilities of an $n$ level quantum system), in particular one may use polar coordinates. Next we establish the ortonormal vectors for particular dimensions:
\begin{enumerate}
\item {\it qubit}: For the qubit one has $n=2$ and  
\[
\vec{p}_e = \frac{1}{2}\left( 1,\, 1\right)\,; \qquad \hat{e}_1 = \frac{1}{\sqrt{2}} \left(1,\,-1\right)\,,
\]
and a point in the simplex in polar representation is given by 
\[
\vec{p} = \vec{p}_e + \frac{1}{\sqrt{2}}\cos(\theta_1)\hat{e}_1\, \, ,
\]
with $0\leq \theta_1 \leq \pi$.
\item {\it qutrit}: In this case $n=3$ 
\[
\vec{p}_e = \frac{1}{3} \left(1,\,1,\,1\right)\,, \quad 
\hat{e}_1 = \frac{1}{\sqrt{2}}\left(1,\,-1,\,0\right)\,, \quad 
\hat{e}_2 = \frac{1}{\sqrt{6}}\left(1,\,1,\,-2\right)\,, 
\]
and a point in the simplex in a polar representation may be written as
\[
\vec{p} = \vec{p}_e + \frac{r}{\sqrt{2}}\left[\cos(\theta_1) \hat{e}_1 + \sin(\theta_1)\hat{e}_2\right]\, \, ,
\] 
with $0\leq \theta_1 \leq 2 \pi$ and $ 0\leq r \leq \frac{2}{\sqrt 3}$.
\item {ququart}: For $n=4$ one has
\begin{eqnarray*}
\vec{p}_e &=& \frac{1}{4}\left(1,\,1,\,1,\,1\right)\,,\qquad  \hat{e}_1 = \frac{1}{\sqrt{2}}\left(1,\,-1,\,0,\,0\right)\,, \\[2mm]
\hat{e}_2 &=& \frac{1}{\sqrt{6}}\left(1,\,1,\,-2,\,0\right)\,,\qquad  \hat{e}_3 = \frac{1}{2\sqrt{3}}\left(1,\,1,\,1,\,-3\right)\,,
\end{eqnarray*}
and the polar representation
\[
\vec{p} = \vec{p}_e + \frac{r}{\sqrt{2}} \left[\cos(\theta_1) \hat{e}_1 + \sin(\theta_1) \cos(\theta_2 )\hat{e}_2 + \sin(\theta_1) \sin(\theta_2) \hat{e}_3 \right]\, ,
\]
with $0\leq \theta_1 \leq \pi$, $0 \leq \theta_2 \leq 2 \pi$ and $0\leq r \leq \sqrt{\frac{3}{2}}$.

\item {qudit}: For the general case one has
\begin{eqnarray*}
\vec{p} = \vec{p}_e + \frac{r}{\sqrt{2}} &&\left[\cos(\theta_1) \hat{e}_1 + \sin(\theta_1) \cos(\theta_2 )\hat{e}_2 + \sin(\theta_1) \sin(\theta_2) \cos(\theta_3) \hat{e}_3 + \,\ldots\, \right.\\
&&\left. + \sin(\theta_1) \sin(\theta_2)\cdots \sin(\theta_{n-1}) \sin(\theta_n) \hat{e}_{n+1} \right]\, .
\end{eqnarray*}
\end{enumerate}

Notice that in general one has that $0\leq r \leq |\vec{\lambda}|_{\rm max}= \sqrt{\frac{2 (n-1)}{n}}$, the magnitude of the Gell-Mann vector, this is so because the boundary of the simplex represents a pure state.

The above expressions allow us to find in a simple way collections of points with a fixed constant invariant value.

\subsection*{Acknowledgments}
J.A. L.-S. wants to acknowledge support from the Priority 2030 program at the National University of Science and Technology “MISIS” under the project K1-2022-027. OC appreciates the support received from PASPA of DGAPA-UNAM (Sep.2023-Aug.2024). EN-A was partially supported by DGAPA-UNAM under project IN100323.

\section*{References}



\begin{thebibliography}{10}
\expandafter\ifx\csname url\endcsname\relax
  \def\url#1{{\tt #1}}\fi
\expandafter\ifx\csname urlprefix\endcsname\relax\def\urlprefix{URL }\fi
\providecommand{\eprint}[2][]{\url{#2}}

\bibitem{bennett98}
Bennett C and Shor P 1998 {\em IEEE Transactions on Information Theory\/} {\bf
  44} 2724--2742

\bibitem{nielsen11}
Nielsen M~A and Chuang I~L 2011 {\em Quantum Computation and Quantum
  Information\/} (Cambridge University Press)

\bibitem{werner01}
Werner R~F 2001 {\em Quantum Information Theory - an Invitation\/} (Berlin,
  Heidelberg: Springer Berlin Heidelberg) pp 14--57 ISBN 978-3-540-44678-1

\bibitem{zoller05}
Zoller P, Beth T, Binosi D, Blatt R, Briegel H, Bruss D, Calarco T, Cirac J~I,
  Deutsch D, Eisert J, Ekert A, Fabre C, Gisin N, Grangiere P, Grassl M,
  Haroche S, Imamoglu A, Karlson A, Kempe J, Kouwenhoven L, {Kr\"oll} S, Leuchs
  G, Lewenstein M, Loss D, {L\"utkenhaus} N, Massar S, Mooij J~E, Plenio M~B,
  Polzik E, Popescu S, Rempe G, Sergienko A, Suter D, Twamley J, Wendin G,
  Werner R, Winter A, Wrachtrup J and Zeilinger A 2005 {\em Eur. Phys. J. D\/}
  {\bf 36} 203--228

\bibitem{lapierre21}
LaPierre R 2021 {\em Introduction to Quantum Computing\/} (Springer Cham)

\bibitem{horodecki22}
Horodecki P, Rudnicki L and Zyczkowski K 2022 {\em PRX Quantum\/} {\bf 3}
  010101

\bibitem{bengtsson17}
Bengtsson I and Zyczkowski K 2017 {\em Geometry of Quantum States\/} 2nd ed
  (Cambridge University Press)

\bibitem{bruss19}
{Bru\ss} D and Leuchs G 2019 {\em Quantum Information: From Foundations to
  Quantum Technology Applications\/} (Wiley-VCH) ISBN 9783527805778

\bibitem{fano57}
Fano U 1957 {\em Rev. Mod. Phys.\/} {\bf 29}(1) 74--93

\bibitem{byrd03}
Byrd M~S and Khaneja N 2003 {\em Phys. Rev. A\/} {\bf 68}(6) 062322

\bibitem{kimura03}
Kimura G 2003 {\em Phys. Lett. A\/} {\bf 314} 339--349 ISSN 0375-9601

\bibitem{akhtarshenas07}
Akhtarshenas S~J 2007 {\em Opt. Spectrosc.\/} {\bf 103} 411--415

\bibitem{bertlmann08}
Bertlmann R~A and Krammer P 2008 {\em J. Phys. A: Mathematical and
  Theoretical\/} {\bf 41} 235303

\bibitem{kurzynski16}
Kurzy\ifmmode~\acute{n}\else \'{n}\fi{}ski P, Ko\l{}odziejski A, Laskowski W
  and Markiewicz M 2016 {\em Phys. Rev. A\/} {\bf 93} 062126

\bibitem{sarbicki12}
Sarbicki G and Bengtsson I 2012 {\em J. Phys. A: Math. Theor.\/} {\bf 46}
  035306

\bibitem{Grunbaum2009-ga}
Grunbaum B 2009 {\em Configurations of Points and Lines\/} Graduate studies in
  mathematics (Providence, RI: American Mathematical Society)

\bibitem{Hirschfeld1986-uz}
Hirschfeld J~W~P 1986 {\em Finite projective spaces of three dimensions\/}
  Oxford Mathematical Monographs (London, England: Oxford University Press)

\bibitem{rau21}
Rau A~R~P 2021 {\em Symmetry\/} {\bf 13} 1732

\bibitem{wang20}
Wang Y, Hu Z, Sanders B~C and Kais S 2020 {\em Front. Phys.\/} {\bf 8} 589504

\bibitem{mandilara24}
Mandilara A, Dellen B, Jaekel U, Valtinos T and Syvridis D 2024 {\em Quantum
  Mach. Intell.\/} {\bf 6} 17

\bibitem{tabia13}
Tabia G~N~M and Appleby D~M 2013 {\em Phys. Rev. A\/} {\bf 88} 012131

\bibitem{xie20}
Xie J, Zhang A, Cao N, Xu H, Zheng K, Poon Y~T, Sze N~S, Xu P, Zeng B and Zhang
  L 2020 {\em Phys. Rev. Lett.\/} {\bf 125} 150401

\bibitem{cordero21}
Cordero S, Nahmad-Achar E, L{\'{o}}pez-Pe{\~{n}}a R and Casta{\~{n}}os O 2021
  {\em Phys. Scr\/} {\bf 96} 035104

\bibitem{lopez-pena21}
L{\'{o}}pez-Pe{\~{n}}a R, Cordero S, Nahmad-Achar E and Casta{\~{n}}os O 2021
  {\em Phys. Scr.\/} {\bf 96} 035103

\bibitem{goyal16}
Goyal S~K, Simon B~N, Singh R and Simon S 2016 {\em J. Phys. A: Math. Theor.\/}
  {\bf 49} 165203

\bibitem{benenti07}
Benenti G, Casati G and Strini G 2007 {\em Principles of Quantum Computation
  and Information\/} vol I, II (World Scientific)

\bibitem{grandy97}
Grandy W~T J 1997 {\em American Journal of Physics\/} {\bf 65} 466--476 ISSN
  0002-9505

\bibitem{shannon48}
Shannon C~E 1948 {\em The Bell System Technical Journal\/} {\bf 27} 379--423

\bibitem{jaynes57a}
Jaynes E~T 1957 {\em Phys. Rev.\/} {\bf 106}(4) 620--630

\bibitem{jaynes57b}
Jaynes E~T 1957 {\em Phys. Rev.\/} {\bf 108}(2) 171--190

\bibitem{louisell90}
Louisell W 1990 {\em Quantum Statistical Properties of Radiation\/} Wiley
  Classics Library (Wiley) ISBN 9780471523659

\bibitem{shore80}
Shore J and Johnson R 1980 {\em IEEE Transactions on Information Theory\/} {\bf
  26} 26--37

\bibitem{jaynes62}
Heims S~P and Jaynes E~T 1962 {\em Rev. Mod. Phys.\/} {\bf 34}(2) 143--165

\bibitem{reichl16}
Reichl L~E 2016 {\em A Modern Course in Statistical Physics\/} 4th ed Physics
  textbook (Wiley) ISBN 978-3-527-41349-2

\bibitem{linden02}
Linden N, Popescu S and Wootters W~K 2002 {\em Phys. Rev. Lett.\/} {\bf 89}(20)
  207901

\bibitem{goldstein06}
Goldstein S, Lebowitz J~L, Tumulka R and Zangh\`{i} N 2006 {\em Phys. Rev.
  Lett.\/} {\bf 96}(5) 050403

\bibitem{tasaki98}
Tasaki H 1998 {\em Phys. Rev. Lett.\/} {\bf 80}(7) 1373--1376

\bibitem{coxeter48}
Coxeter H~S~M 1948 {\em Regular Polytopes\/} (Methuen \& Co. Ltd., London)

\bibitem{gilmore05}
Gilmore R 2005 {\em Lie Groups, Lie Algebras, and Some of Their Applications\/}
  Dover Books on Mathematics (Dover Publications) ISBN 9780486131566

\bibitem{petruccione12}
Br\"uning E, M\"akel\"a H, Messina A and Petruccione F 2012 {\em J. Modern
  Optics\/} {\bf 59} 1--20

\bibitem{LMG65a}
Lipkin H~J, Meshkov N and Glick A~J 1965 {\em Nucl. Phys.\/} {\bf 62} 188

\bibitem{LMG65b}
Meshkov N, Glick A~J and Lipkin H~J 1965 {\em Nucl. Phys.\/} {\bf 62} 199

\bibitem{LMG65c}
Glick A~J, Lipkin H~J and Meshkov N 1965 {\em Nucl. Phys.\/} {\bf 62} 211

\bibitem{botet82}
Botet R, Jullien R and Pfeuty P 1982 {\em Phys. Rev. Lett.\/} {\bf 49}(7)
  478--481

\bibitem{botet83}
Botet R and Jullien R 1983 {\em Phys. Rev. B\/} {\bf 28}(7) 3955--3967

\bibitem{citlali10}
{P\'erez-Campos} C, {Gonz\'alez-Alonso} J~R, Casta\~nos O and R L 2010 {\em
  Annals of Physics\/} {\bf 325} 325--344

\bibitem{kitagawa93}
Kitagawa M and Ueda M 1993 {\em Phys. Rev. A\/} {\bf 47}(6) 5138--5143

\bibitem{calixto21}
Calixto M, Mayorgas A and Guerrero J 2021 {\em Quantum Information
  Processing\/} {\bf 20} 304

\bibitem{guerrero22}
Guerrero J, Mayorgas A and Calixto M 2022 {\em Quantum Information
  Processing\/} {\bf 21} 223

\bibitem{mayorgas23}
Mayorgas A, Guerrero J and Calixto M 2023 {\em Phys. Rev. E\/} {\bf 108}(2)
  024107

\bibitem{procesi78}
Procesi C 1978 {\em Advances in Mathematics\/} {\bf 29} 219--225 ISSN 0001-8708

\bibitem{procesi07}
Procesi C 2007 {\em Lie Groups: An approach through Invariants and
  representations\/} (Springer New York, NY)

\bibitem{curtiss18}
Curtiss D~R 1918 {\em Annals of Mathematics\/} {\bf 19} 251--278 ISSN 0003486X

\bibitem{arvind97}
Arvind, Mallesh K~S and Mukunda N 1997 {\em J. Phys. A: Mathematical and
  General\/} {\bf 30} 2417

\bibitem{khanna97}
Khanna G, Mukhopadhyay S, Simon R and Mukunda N 1997 {\em Ann. of Phys.\/} {\bf
  253} 55--82 ISSN 0003-4916

\bibitem{kus01}
Ku\ifmmode~\acute{s}\else \'{s}\fi{} M and \ifmmode~\dot{Z}\else
  \.{Z}\fi{}yczkowski K 2001 {\em Phys. Rev. A\/} {\bf 63}(3) 032307

\bibitem{bengtsson06}
Bengtsson I and Zyczkowski K 2006 {\em Geometry of quantum states: An
  introduction to quantum entanglement\/} (Cambridge University Press)

\bibitem{tay08}
Tay B~A and Zainuddin H 2008 {\em Chinese Phys. Lett.\/} {\bf 25} 1923

\bibitem{gerdt14}
Gerdt V~P, Khvedelidze A~M and Palii Y~G 2014 {\em J. of Math. Sciences\/} {\bf
  200} 682--689

\bibitem{bruning12}
{Br\"uning} E, {M\"akel\"a} H, Messina A and Petruccione F 2012 {\em J. of
  Modern Optics\/} {\bf 59} 1

\end{thebibliography}

\providecommand{\newblock}{}

\end{document}